\documentclass[twocolumn]{aastex63}

\usepackage{graphicx, amsmath}
\usepackage[caption=false]{subfig}
\usepackage{hyperref}
\usepackage{cleveref}
\usepackage{xspace}
\usepackage{rotating}
\usepackage{relsize}
\usepackage{comment}

\newcommand{\lya}{Ly$\alpha$\xspace}

\newcommand{\Ha}{H$\alpha$\xspace}

\newcommand{\angstrom}{\mbox{\normalfont\AA\xspace}}
\newcommand{\OII}{[\ion{O}{2}]\xspace}
\newcommand{\OIII}{[\ion{O}{3}]\xspace}

\newcommand{\MgII}{[\ion{Mg}{2}]\xspace}
\newcommand{\CII}{[\ion{C}{2}]\xspace}
\newcommand{\CIII}{[\ion{C}{3}]\xspace}
\newcommand{\CIV}{[\ion{C}{4}]\xspace}

\newcommand{\oii}{[O\,{\sc ii}]}  
\newcommand{\ovi}{[O\,{\sc vi}]}

\usepackage{enumitem,amssymb}
\usepackage{float}
\usepackage[utf8]{inputenc}
\usepackage{color}
\usepackage{graphicx} 
\usepackage{amsmath} 
\usepackage{amssymb} 
\usepackage{wasysym}
\usepackage{mathtools}
\usepackage{newtxtext,newtxmath}
\def\arcs{\hbox{$^{\prime\prime}$}}

\newcommand{\cgsa}{erg s$^{-1}$ cm$^{-2}$ $\angstrom^{-1}$}
\newcommand{\cgs}{erg s$^{-1}$ cm$^{-2}$}

\newcommand{\lycujy}{0.10 $\pm$0.036 $\mu$Jy}

\newcommand{\lyccount}{214}

\graphicspath{{./}{figures/}}

\let\textbf\relax

\shorttitle{Lyman Continuum in HETDEX}
\shortauthors{Davis et al}

\begin{document}

\title{Detection of Lyman Continuum from 3.0 < $z$ < 3.5 Galaxies in the HETDEX Survey}

\author[0000-0002-8925-9769]{Dustin Davis}
\affiliation{Department of Astronomy, The University of Texas at Austin, 2515 Speedway Boulevard, Austin, TX 78712, USA}

%
%
\author[0000-0002-8433-8185]{Karl Gebhardt}
\affiliation{Department of Astronomy, The University of Texas at Austin, 2515 Speedway Boulevard, Austin, TX 78712, USA}

\author[0000-0002-2307-0146]{Erin Mentuch Cooper}
\affiliation{Department of Astronomy, The University of Texas at Austin, 2515 Speedway Boulevard, Austin, TX 78712, USA}

\author[0000-0002-0302-2577]{John Chisholm}\altaffiliation {Hubble Fellow}
\affiliation{Department of Astronomy, The University of Texas at Austin, 2515 Speedway Boulevard, Austin, TX 78712, USA}

\author[0000-0002-1328-0211]{Robin Ciardullo} \affiliation{Department of Astronomy \& Astrophysics, The Pennsylvania
State University, University Park, PA 16802, USA}
\affiliation{Institute for Gravitation and the Cosmos, The Pennsylvania State University, University Park, PA 16802, USA}

\author{Daniel J. Farrow}
\affiliation{Max-Planck Institut f\"ur extraterrestrische Physik, Giessenbachstrasse 1, 85748 Garching, Germany}
\affiliation{Universit{\"a}ts-Sternwarte, Fakult{\"a}t f{\"u}r Physik, Ludwig-Maximilians Universit{\"a}t M{\"u}nchen, Scheinerstr. 1, 81679\ M{\"u}nchen, Germany}

\author[0000-0001-8519-1130]{Steven L. Finkelstein}
\affiliation{Department of Astronomy, The University of Texas at Austin, 2515 Speedway Boulevard, Austin, TX 78712, USA}

\author[0000-0001-6842-2371]{Caryl Gronwall}
\affiliation{Department of Astronomy \& Astrophysics, The Pennsylvania
State University, University Park, PA 16802, USA}
\affiliation{Institute for Gravitation and the Cosmos, The Pennsylvania State University, University Park, PA 16802, USA}

\author[0000-0003-1530-8713]{Eric Gawiser}
\affiliation{Department of Physics \& Astronomy, Rutgers, The State University of New Jersey, Piscataway, NJ 08854, USA}

\author{Gary J. Hill}
\affiliation{Department of Astronomy, The University of Texas at Austin, 2515 Speedway Boulevard, Austin, TX 78712, USA}
\affiliation{McDonald Observatory, The University of Texas at Austin, 2515 Speedway Boulevard, Austin, TX 78712, USA}

\author[0000-0003-1008-225X]{Ulrich Hopp}
\affiliation{Universit{\"a}ts-Sternwarte, Fakult{\"a}t f{\"u}r Physik, Ludwig-Maximilians Universit{\"a}t M{\"u}nchen, Scheinerstr. 1, 81679\ M{\"u}nchen, Germany}
\affiliation{Max-Planck Institut f\"ur extraterrestrische Physik, Giessenbachstrasse 1, 85748 Garching, Germany}

\author[0000-0002-8434-979X]{Donghui Jeong}
\affiliation{Department of Astronomy \& Astrophysics, The Pennsylvania State University, University Park, PA 16802, USA}
\affiliation{Institute for Gravitation and the Cosmos, The Pennsylvania State University, University Park, PA 16802, USA}

\author{Martin Landriau}
\affiliation{Lawrence Berkeley National Laboratory, 1 Cyclotron Road, Berkeley, CA 94720, USA}

\author[0000-0001-5561-2010]{Chenxu Liu}
\affiliation{Department of Astronomy, The University of Texas at Austin, 2515 Speedway Boulevard, Austin, TX 78712, USA}

\author{Maja Lujan Niemeyer}
\affiliation{Max-Planck-Institut f\"ur Astrophysik, Karl-Schwarzschild-Str. 1, 85748 Garching, Germany}

\author{Donald P. Schneider}
\affiliation{Department of Astronomy \& Astrophysics, The Pennsylvania State University, University Park, PA 16802, USA}
\affiliation{Institute for Gravitation and the Cosmos, The Pennsylvania State University, University Park, PA 16802, USA}

\author[0000-0003-4044-5357]{Jan Snigula}
\affiliation{Universit{\"a}ts-Sternwarte, Fakult{\"a}t f{\"u}r Physik, Ludwig-Maximilians Universit{\"a}t M{\"u}nchen, Scheinerstr. 1, 81679\ M{\"u}nchen, Germany}
\affiliation{Max-Planck Institut f\"ur extraterrestrische Physik, Giessenbachstrasse 1, 85748 Garching, Germany}

\author[0000-0002-7327-565X]{Sarah Tuttle}
\affiliation{Department of Astronomy, University of Washington, Seattle, 3910 15th Ave NE, Room C319, Seattle WA 98195-0002}

\defcitealias{Gebhardt+2021}{KG21}
\defcitealias{Davis2021}{DD21}

\begin{abstract}

Questions as to what drove the bulk reionization of the Universe, how that reionization proceeded, \textbf{and how the hard ionizing radiation reached the intergalactic medium} remain open and debated. Observations probing that epoch are severely hampered by the increasing amounts of neutral gas with increasing redshift, so a small, but growing number of experiments are targeting star forming galaxies ($z\sim3$) as proxies. However, these studies, while providing fantastic detail, are time intensive, contain relatively few targets, and can suffer from selection biases. As a complementary alternative, we investigate whether stacking the already vast (and growing) numbers of low-resolution ($\Delta \lambda / \lambda = 800$) Lyman-$\alpha$ Emitting (LAE) galaxy spectra from the Hobby-Eberly Telescope Dark Energy Experiment (HETDEX) can be used to measure ionizing photons (restframe 880-910\AA) escaping their galaxy hosts. As a blind survey, HETDEX avoids the biases from continuum selected galaxies and its planned 540 square degree coverage promotes the statistical power of large numbers. In this paper, we confirm the feasibility of Lyman continuum detection by carefully selecting a sample of \lyccount\ high redshift ($z\sim$3) LAEs from a subset of HETDEX observations, stacking their spectra and measuring a $\gtrsim$3$\sigma$ detection of $0.10 \mu$Jy restframe Lyman continuum emission, uncorrected for attenuation in the intergalactic medium, over the full sample stack ($3.0 < z < 3.5$ and $-22.0 \lesssim M_{\text{UV}} \lesssim -19.0$). 

\end{abstract}


\section{Introduction}\label{sec:intro} 
\textbf{As the universe began to assemble galaxies from z$\sim$15 and onward \citep{Bromm_2011}}, Extreme Ultra-Violet (EUV) or Lyman continuum photons (here, defined specifically as those photons near but shortward of 912~\AA) began filling the Universe, ionizing the bulk, neutral gas between the galaxies so that by $5 \lesssim z \lesssim 8$ the intergalactic medium (IGM) was once again (almost) entirely ionized \citep{Finkelstein_2016, Stark,becker2021}. This period that witnesses this change of state in the hydrogen gas of the IGM, roughly $5 \lesssim z \lesssim 15$ (though the beginning redshift is rather more uncertain than the end), is commonly called the Epoch of Reionization (EoR) (\citet{Gunn_Peterson,Finkelstein_2019}, and many others).   

While it is well accepted that the driver of this reionization is photoionization, and that O-type stars and Active Galactic Nuclei (AGN)/quasars are prodigious producers of the EUV photons necessary to completely ionize hydrogen, the large-scale mechanics and evolution of the EoR are less well understood. Simulations and limited observations suggest a ``Swiss-cheese'' model, with bubbles of ionized gas growing and merging into larger and larger regions until only small pockets of neutral gas remain \citep{Zaroubi_2012}. Almost certainly the young galaxies are the origin of these extra-galactic bubbles in which they reside, \textbf{but whether the principal actors are massive stars in (1) small, young star forming galaxies, (2) older, more massive galaxies, (3) accreting Super Massive Black Holes (SMBH) in the galaxy centers, or (4) some other source is unclear.}

Ideally, this issue could be resolved empirically by counting the Lyman continuum photons escaping from various galaxy hosts during the EoR, but these photons are unavailable to be counted as they are consumed in the reionization of the intergalactic hydrogen. Consequently, we need to observe galaxies near the EoR but after reionization is complete. Unfortunately, as we approach those higher redshifts, increasing numbers of remnant clumps of neutral hydrogen cloud our lines of sight such that by $z\gtrsim$4 the Lyman continuum photons are effectively all absorbed \citep{Haardt1995,Meiksin,Cowie1998,Overzier,Vanzella_2018}. If, however, we look at galaxies near redshift $\sim$3, some of the ionizing photons escaping their hosts survive to reach our telescopes. We can then use these lower redshift galaxies as analogs of their EoR predecessors, matching as closely as possible against various properties such as halo mass, stellar mass, metallicity, stellar population, and star formation rate (\citet{Shapley_2003, Shapley_2006,Vanzella_2015, Shapley_2016, Steidel_2018}, and others). 

The deep spectroscopic observations needed to measure these photons from individual galaxies are difficult and time consuming in this redshift range, \textbf{so pre-selection of targets is necessary. This pre-selection is typically performed with a filter drop out in broadband photometric imaging, bracketing the Lyman Break (and the so named, Lyman Break Galaxies or LBGs) \citep[][]{Steidel_1993,Steidel_1996,Dickinson_1996}. This is a robust technique, but not without issues, as is discussed in Section \ref{sec:discussion}, and there is no guarantee that a targeted galaxy will be in a clear sight line (generally devoid of dense clumps of HI).} There is also the real risk, particularly without deep broadband imaging accompanying the spectra, that what is thought to be Lyman continuum is actually from lower redshift interlopers \citep{Vanzella_2010,Vanzella_2012,Steidel_2018,pahl_2021}. \textbf{These galaxies tend to be fairly evolved and, as a consequence of the selection technique, moderately large and continuum bright.}

\textbf{On the other hand, Lyman-$\alpha$ Emitting (LAE) galaxies tend to be less massive, UV-fainter and more difficult to detect in broadband imaging than their LBG siblings. They appear to increase as a fraction of the overall population with increasing redshift \citep[][]{Stark_2010,Kornei_2010,finkelstein_2010,Trenti_2010,Jose_2013,Naidu_2020,ito_2021,santos_2021}, making them more representative of EoR galaxies than the z$\sim$3 LBGs. As such, they represent an important sub-population in the understanding of galaxy evolution and of Lyman continuum escape, while simultaneously being (historically) more difficult to observe as spectroscopic identification is necessary.} \textbf{However,} blanketing large areas of sky with low-cost, fast observations (meaning, shallow, low-resolution spectra), would vastly increase the number of sampled \textbf{LAEs}. Even though Lyman continuum emission would rarely be detected above the noise for individual galaxies, we can stack the spectra, marginalizing over sight lines and galaxy orientations, to boost the signal.

The Hobby-Eberly Telescope Dark Energy Experiment (HETDEX) \citep{Hill+2008, Hill+21,Gebhardt+2021} is a massive, multi-year blind spectroscopic survey of some 540 square degrees of sky. HETDEX employs the Visible Integral-field Replicable Unit Spectrograph (VIRUS) on the upgraded 10 m Hobby-Eberly Telescope (HET). VIRUS has up to 74 pairs of integral field spectrographs fed by more than 30,000 fibers \citep{Hill+21}. HETDEX is mapping out the 3D positions in Right Ascension, Declination, and redshift ($\alpha,\delta,z$) of $\sim 10^6$ Lyman-$\alpha$ Emitting (LAE) galaxies (detected due to their bright emission in Ly$\alpha$) approximately $z$ $\in$ (1.9,3.5) as well as similar number of low redshift ($z$ $\in$ (0.0,0.5)) galaxies. The low-z galaxies are detected mostly in \OII 3727~\AA\ and to a lesser extent, \OIII 4959, 5007~\AA, and the Balmer lines (excluding \Ha), as well as mid-range redshifts ($z$ $\in$ (0.7,1.9)) with \MgII 2795+2802~\AA, \CIII 1909~\AA, and \CIV 1549~\AA. 

In August 2020, HETDEX completed its second updated dataset (designated iHDR2) with roughly 1.5 million emission line detections from the first three years of observations. Using the detections from the Spring 2020 observations with higher ($>$6) emission line signal-to-noise ratios (SNR), as defined in \citet[][hereafter KG21]{Gebhardt+2021}, we select galaxies with Ly$\alpha$ emission at the higher-redshift end (3.0 $< z < $3.5) of the survey, where the Lyman continuum region falls within the HETDEX spectral range. The dataset is further refined to include only those LAEs that are, as best we can determine, free from any significant contamination from lower $z$ foreground objects (see Section \ref{sec:selection}). We then stack the sample spectra to detect Lyman continuum emission. 

\textbf{
This feasibility study suggests the massing of large samples of LAEs, particularly z$\gtrsim$3, is now practical. This enables the investigation of the ensemble properties of this subset of the galaxy population, farther down the galaxy mass function which may, by the virtue of their increasing numbers and possibly enhanced Lyman continuum leakage \citep{Yan_2004,bouwens_2015,finkelstein_2015,Finkelstein_2019}, be from the class of objects largely responsible for the ionizing photon budget of Reionization. Translating the observations of these z$\sim$3 LAEs to galaxies at z$\gtrsim$6 will be addressed in work and will be approached via the mapping of similar properties (halo mass, stellar mass, star formation rates, etc) onto modeled galaxies for the EoR, correcting for bias and number density evolution. This also holds the promise of an enormous catalog of galaxies, specially curated as a test bed to study the physics of how EUV photons escape into the IGM.
}


The paper is arranged as follows: Section \ref{sec:hetdex} provides an overview of the HETDEX survey and its data processing pipeline. Section \ref{sec:selection} describes the critical steps taken to assure a pure sample of 3.0 $<$ $z$ $<$ 3.5 LAE galaxies, free from foreground contamination. Section \ref{sec:analysis} reviews the data analysis and stacking methods. Section 5 \textbf{highlights a limited comparison to a similar study and discusses various biases and issues.} Section 6 presents a brief summary with the conclusion that restframe Lyman continuum flux has been detected with stacked HETDEX observations and previews future work.

Throughout the paper, a concordance (flat-$\Lambda$CDM) cosmology with $\Omega_{\Lambda}$= 0.7,  $\Omega_{\text{m}}$= 0.3 and H$_0$ = 70 $\mathrm{km~s^{-1}~Mpc^{-1}}$ is assumed. All magnitudes are in the AB system.\\


\section{Observations} \label{sec:hetdex}

The Point Spread Function (PSF) weighted, extracted spectra and astrometric solutions from HETDEX are taken as the initial, primary inputs to this work. An extensive discussion of the HETDEX project (science goals, design and instrumentation, data reduction pipeline, modelling and calibration, emission line detection, and more) is presented in \citetalias{Gebhardt+2021} and \citet{Hill+21}. Briefly, each HETDEX observation consists of three 360s exposures with VIRUS, offset in a triangular dither pattern to fill in the spatial gaps in the fiber layout of the (up to) 74 Integral Field Unit (IFU) spectrographs. Each VIRUS IFU covers 50x50 square arcseconds area with 448, 1$\farcs$5 diameter fibers arranged in a hexagonal grid pattern with 1$/$3 fill factor on the IFU. Each IFU feeds light from the prime focus of the 10m Hobby-Eberly telescope to a pair of spectrographs covering $\sim$3500 - 5500~\AA\ with $\Delta \lambda / \lambda = 800$. The IFUs are arranged in a square grid on 100 arcsecond centers. Thus, each exposure contains more than $3\times 10^{4}$ low-resolution spectra and an observation of three dithered exposures covers more than 50 square arcminutes sky area spread over an 18' diameter field of view with a 1:4.6 fill factor.

While the most recent HETDEX Data Release (version 2.1, at the time of this writing) contains 3266 observations (each as 3 dithers per pointing) from January 2017 through June 2020, only the 816 observations within the first half of 2020 (January through June) and inside the primary HETDEX spring field (roughly RA 160$^{\circ}$ to 245$^{\circ}$ and Dec 45$^{\circ}$ to 57$^{\circ}$) are included for consideration in the current work. This set of observations is further reduced to 662 by enforcing reasonable throughput (larger than 0.095 at 4540~$\angstrom$; the sample median is above 0.120) and seeing FWHM (below 2$\farcs$6; with the sample median better than 1$\farcs$7) \citepalias[see][for details]{Gebhardt+2021}. This restricted base set of observations is selected to focus on regions where deep, supplemental photometric imaging is available as well as to expedite the review and classification process (see Section \ref{sec:selection}), which includes a lengthy visual inspection component. All LAEs included in this work have deep $r$ ($\sim$26) coverage from an extensive Subaru Hyper Suprime-Cam (HSC) survey specially executed in support of HETDEX. A detailed description and analysis of this and other photometric surveys used in HETDEX is provided in \citet[][hereafter DD21]{Davis2021}. 

Reported $r$ magnitudes are from variable (typically 0\farcs5-1\farcs0 semi-major axis) elliptical apertures \citep[][]{source_extractor,sep_pkg} placed on the HSC imaging. Where no counterpart is found or the aperture magnitude is negative or fainter than the limiting magnitude, that limit is interpreted as the magnitude. For HSC-$r$, the $5\sigma$ limiting magnitude is $\sim26.5$ ($\sim25.5$) for 1\farcs0 (2\farcs0) diameter circular aperture. As the seeing FWHM in this HSC survey varies between 0\farcs6 and 1\farcs0 and the LAEs of this work are point-source like, 26.5 is adopted as the limiting magnitude. 

The HSC $r$-band uses the SDSS-$r$ filter \citep{Doi+2010} which conveniently places the rest-frame 1500~\AA\ flux squarely within HSC-$r$, while avoiding the \lya\ line for the redshift range of this work, allowing the simple use of the distance modulus adjusted $r$-magnitude as the absolute restframe UV magnitude.
\begin{equation} \label{eq:r_muv}
    M_{\text{UV}} = r - 5\log( \frac{D_\text{L}}{10\text{pc}}) + 2.5\log(1+z_{\text{Ly}\alpha}) + K,
\end{equation}
where $D_\text{L}$ is the luminosity distance in parsecs and $K$ is the K-correction, which is set to 0 as $r$ probes the UV region for the redshift range of this paper. The $2.5\log(1+z_{\text{Ly}\alpha})$ term for the band-pass compression is often included in the $K$ term, but is separated out here for clarity.

\section{Data Selection} \label{sec:selection}

Prior to consumption in this work, the data are reduced (producing sky-subtracted, wavelength rectified, flux-calibrated spectra for each fiber) and initial detections are made, combining multiple fibers inside apertures and weighting according to a point source model. Details of the reduction pipeline and detection methods can be found in \citetalias{Gebhardt+2021}.

The principal investigative tool for detections is the HETDEX Emission Line eXplorer (ELiXer), which combines the HETDEX reduced data and archival imaging surveys into reports for each emission line detection to aid line identification and observation diagnostics. It performs automated and semi-automated characterization and classifications, using multiple techniques and information sources, including Bayesian based Ly$\alpha$ vs \OII $\lambda$ 3727 analysis \citep{Leung,Farrow+2021}, photometric catalog matching, emission line groups and flux ratios, redshift corrected physical extents, etc.  The ELiXer software is described in detail in \citetalias{Davis2021}.

As it is absolutely critical that the data sample be as free as possible of misidentifications and from faint continuum interlopers along or near the line of sight, strict selection criteria are applied. Each LAE in this sample is manually vetted and is, within the spatial resolution limits of the overlapping imaging, required to be effectively isolated -- meaning without any significant overlap in PSF from other sources. As such, the selection criteria are chosen not only to isolate a clean sample (of LAEs), but also to keep the number of detections manageable for the manual examination. The full iHDR2 catalog contains approximately $1.5\times 10^{6}$ individually fitted (\citetalias{Gebhardt+2021}) emission line entries belonging to a few $\times 10^{5}$ unique astrophysical objects, including well over $1\times 10^{5}$ LAEs. Since the final sample is down-selected to include only \lyccount\ galaxies, certainly most valid LAEs are left "on the table" to be used in future work with more sophisticated contamination mitigation. 

Figure \ref{fig:summary_stats} presents some basic summary statistics for the final sample. Individual panels are also referenced throughout this work, but in brief, from left to right and top to bottom the panels show the sample distributions of:
(\textit{Panel 1}) the signal-to-noise ratio of the Ly$\alpha$ emission line, (\textit{Panel 2}) the redshift, (\textit{Panel 3}) the rest-frame Ly$\alpha$ equivalent width (using the estimated continuum from the HETDEX spectra), (\textit{Panel 4}) the Ly$\alpha$ line luminosity, (\textit{Panel 5}) the $r$ apparent magnitude, and (\textit{Panel 6}) the absolute UV magnitude. Where appropriate, the fractions of the histogram bins with candidate LAEs below the imaging survey's magnitude limit (that is, without an imaging counterpart or with an aperture magnitude fainter than the nominal limit) are marked with red hashes. The Ly$\alpha$ line luminosity simply uses the definition of luminosity,
\begin{equation} \label{eq:lum}
    L_{\text{Ly}\alpha} = 4\pi F D_{\text{L}}^2,
\end{equation}
where $F$ is the integrated line flux and $D_{\text{L}}$ is the luminosity distance.

The initial sample selection criteria seek to eliminate obvious contaminants, spurious detections, and undesirable (to this work) galaxies (specifically AGN with obvious continuum or broad lines, potentially interacting galaxies, or LAEs near bright foreground objects). HETDEX defines its own specifications on the acceptable contamination rate and implements its own procedures to ensure the specifications are met \citepalias{Gebhardt+2021}. However, where the principle science objectives of HETDEX rely primarily on accurate 3D-space mappings ($\alpha,\delta,z$) and is generally less concerned with the LAE galaxy type (AGN, interacting, etc) or whether there is some contamination from the scattered light of nearby objects (so long as it does not affect the redshift determination), this work requires additional refinement to eliminate these contaminants. The first step in reducing the potential LAE candidates from the set of all candidate detections in iHDR2 is a simple threshold filter over several criteria, described in the following subsections. Although the selection criteria are necessarily presented in a sequence below, except where noted, \textbf{the actual execution is as a compound query executed as a single statement.}\\

\subsection{Initial Emission Line Database Selection} \label{subsec:initial_selection}
\begin{figure*}[ht]
    \centering
    \includegraphics[width=0.89 \textwidth]{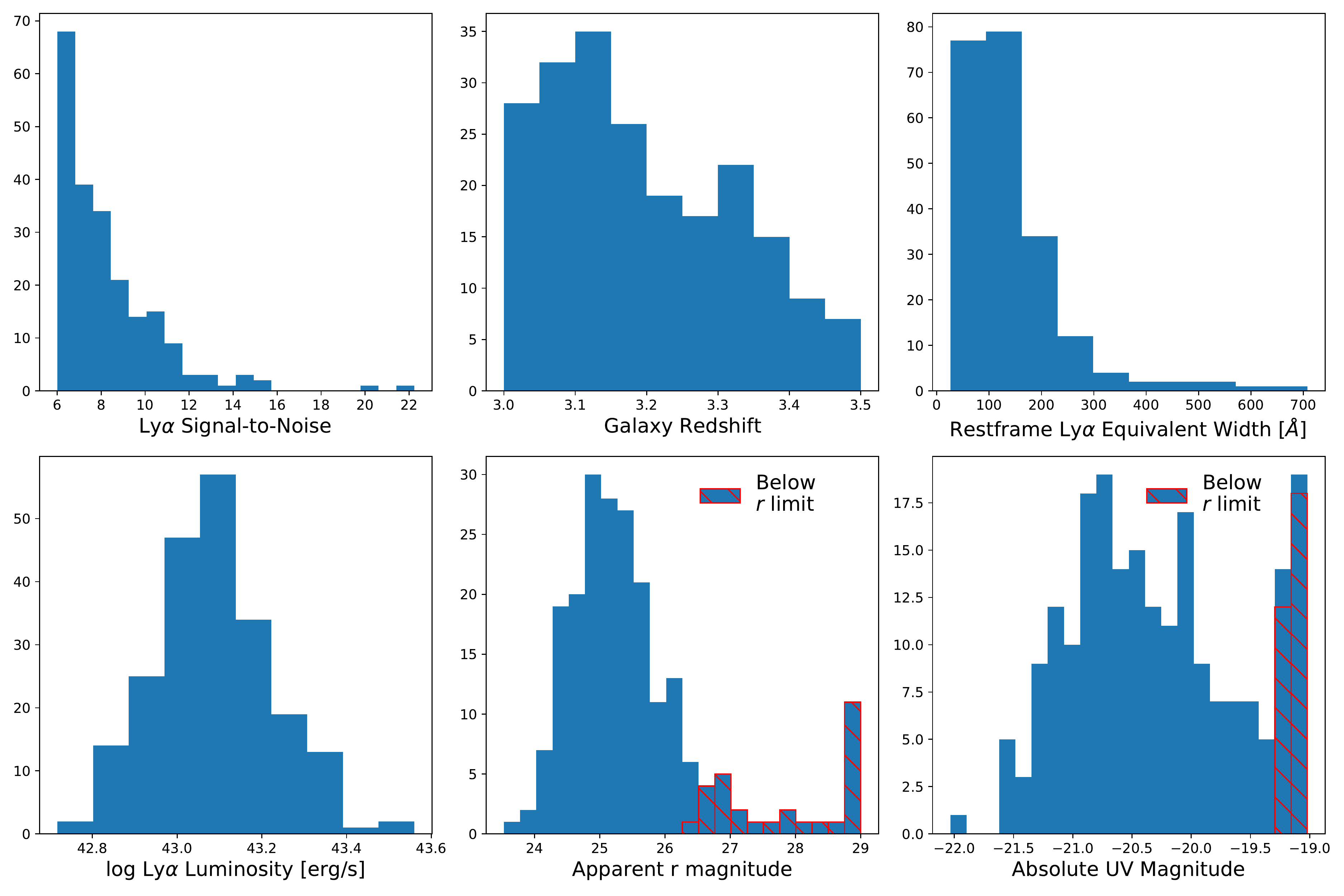}
    \caption{  
        Panels are enumerated from left to right then top to bottom beginning in the upper-left.
        \textit{Panel 1.} Histogram of Ly$\alpha$ SNR for this work. The trend to increasing LAE count with lower SNR is expected to continue down below SNR < 5, where they begin becoming statistically indistinguishable from noise. 
        \textit{Panel 2.} Redshift histogram of final sample (3 $< z <$ 3.5). The limits are selected so that the Lyman continuum region and the Ly$\alpha$ line fall cleanly within the HETDEX spectral range. 
        \textit{Panel 3.} The Ly$\alpha$ restframe equivalent width (EW) distribution of this sample estimated from the HETDEX spectra. Because this estimate includes wavelengths blueward of Ly$\alpha$ there may be extra IGM attenuation and the true EW is likely slightly lower.
        \textit{Panel 4.}  \lya\ line luminosity distribution of this sample, with a median near 1.2$\times$10$^{43}$ erg/s.
        \textit{Panel 5.} The estimated apparent magnitudes from aperture photometry on HSC-$r$. The HSC-$r$ imaging has a 5$\sigma$ limiting depth of $\sim$26.5 mag. The $r$ magnitude was used only in classification steps and the specific value is increasingly unimportant fainter than 25 \citepalias{Davis2021}.
        \textit{Panel 6.} The estimated UV absolute magnitude (uncorrected for IGM attenuation, etc).
        }
    \label{fig:summary_stats}
\end{figure*}

\begin{enumerate}

\item \textit{Emission Wavelength}: 
Since the spectral region of interest is in the Lyman continuum (880-910~\AA) and the HETDEX observable wavelength range is 3470-5540~\AA, the minimum observed emission line wavelength is selected as 4860~\AA\ (or $z$ $\approx$ 3 for Ly$\alpha$, see panel 2 in Figure \ref{fig:summary_stats} and Figure \ref{fig:example_sky_spec}). This ensures the blue end of our targeted Lyman continuum range is captured by the VIRUS spectrographs allowing for some small variation in each detector's wavelength range and avoiding the edge-most pixels. The wavelength limits for the Lyman continuum are chosen to select the photons responsible for reionization (those blue-ward of the Lyman Limit) in the region surrounding the host galaxies without excessive contamination from outside sources. The (red) upper bound is fixed by the Lyman Limit itself while the (blue) lower limit is a bit softer and is established by the mean free path of Lyman continuum photons in 3 $\leq z \leq$ 4. The optical depth of the IGM at $z$ $\sim$ 3 results in a mean free path of $\Delta z$ $\approx$ 0.18, which in turn sets the blue limit to $\sim$870~\AA\ at $z$ = 3 and $\sim$880~\AA\ by $z$ = 4 \citep{Haardt1998, Rudie_2012}. Since the 880-910~\AA\ range is commonly used in other publications, it is adopted here as well for consistency.

\item \textit{Emission Line FWHM and $g$-band Apparent Magnitude}: 
To exclude AGN and brighter contaminants, the maximum allowed fit FWHM to the emission line is set at 600 km s$^{-1}$ with a bright $g$-band magnitude limit of 23. While AGN in the redshift range of interest are certainly Lyman-$\alpha$ Emitters, in this work we are interested in the far more common, simple star forming galaxies (though AGN and broadline/bright LAEs will be included in future work). This selection criteria does not absolutely guarantee the complete exclusion of AGN (particularly faint, narrow line Type-II AGN), but the 600 km s$^{-1}$ limit is well below the more typical $\sim$1000 - 2000 km s$^{-1}$ used in classifying Type-I (broadline) AGN (\citet{Antonucci_1993,Urry_1995,Steidel_2002,Baron_2016}, and others). Future work will explore the possible contamination by Type-II AGN, but here the impact is assumed to be negligible. The photometric $g$ coverage for this sample is incomplete and the $g$ magnitude used here is computed by passing the HETDEX collected spectrum through the SDSS-$g$ filter \citep{Doi+2010} using the Python \textit{speclite} package (version 0.8; \citet{speclite_pkg}). The estimated apparent $g$ magnitude is limited by the HETDEX flux sensitivity to 24.5-25.0 (depending on the observing conditions, exposure time, and the VIRUS IFU) \citepalias{Gebhardt+2021} and most of the band-pass related computation is based on additional photometric imaging (generally $r$ for this work).

\item \textit{Signal-to-Noise Ratio and $\chi^{2}$ Model Fit}: 
Although there is good recovery (at the time of this writing, the detailed HETDEX survey recovery rate at the lower SNR range is still being evaluated) down to a SNR of approximately 5 for the Ly$\alpha$ emission line, and the absolute number of LAEs grows rapidly with decreasing SNR (Figure \ref{fig:summary_stats} panel 1), the relative number of false detections also increases with decreasing SNR and, in an effort to keep this preliminary manual examination sample manageable, an emission line SNR minimum threshold of 6.0 is enforced.  Closely related to the SNR cut, a $\chi^{2}$ limit of 1.2 on the single Gaussian emission line fit is imposed to remove potentially questionable detections. While there certainly are real sources with a poor $\chi^{2}$ fit (particularly broad, noisy, or strongly asymmetric lines) this criterion performs generally well. This selection, combined with the redshift range and emission line FWHM cuts, result in a moderately bright  Ly$\alpha$ line luminosity distribution centered near $1.2\times 10^{43}\ \text{erg}\ \text{s}^{-1}$ (panel 4 in Figure \ref{fig:summary_stats}).

\item \textit{P(Ly$\alpha$) Classification}: \label{sec:plya_classification}
As an obvious condition, candidates that are not strongly indicated as LAEs are excluded from the initial selection with a minimum required value of 0.8 on the ELiXer composite P(Ly$\alpha$) classification value. While expressed as a real number (0-1), P(Ly$\alpha$) is not a proper probability but does represent a form of confidence in the classification of the emission line as Ly$\alpha$. The full details of the computation are presented in \citetalias{Davis2021}. In brief, however, it is a weighted combination of an analysis incorporating several equivalent width estimates, redshifts, and line fluxes \citep{Farrow+2021}, based on the Bayesian analysis in \citet{Leung}. Also factoring in are emission line positions, flux ratios, and other estimated physical and spectral properties from the HETDEX spectra and available photometry. 

The Ly$\alpha$ equivalent width is estimated as 
\begin{equation} \label{eq:EW}
    EW = \frac{F}{C(1+z_{\text{Ly}\alpha})},
\end{equation}
where $F$ is the integrated line flux as fit by HETDEX \citepalias{Gebhardt+2021} and $C$ is the continuum flux density as estimated, in this work, from either the $r$ aperture magnitude or the SDSS-$g$ magnitude from the HETDEX spectrum, as described earlier. While the Ly$\alpha$ line appears within $g$ for the redshifts of this work, $g$-band photometry is often not available or is not as deep as $r$ and so the $r$-magnitude is used, assuming a flat continuum between Ly$\alpha$ and $r$, as the primary continuum estimate in the equivalent width calculation. This does not alter the efficacy of the equivalent width analysis on which the emission line discrimination is based and, in fact, may produce a cleaner separation \citep{Adams_2011,Leung}. 

The Ly$\alpha$ EW from the HETDEX spectra is more similar to a proper equivalent width in that it estimates the continuum nearer the emission line (as opposed to the $r$ magnitude). The sample distribution of this EW estimate is show in panel 3 of Figure \ref{fig:summary_stats} with the caution that the continuum estimates include wavelengths blueward of Ly$\alpha$, subject to increased IGM attenuation, as well as the Ly$\alpha$ line itself. These two issues at least partly cancel, but are sources of error in the EW estimates. Future work will refine the EW estimates and the current results are shown here only in the context of the broad description of the dataset. With that caveat, the smallest rest-frame emission line equivalent width in the final dataset is $\sim$40~\AA\ for the $r$ continuum estimate and $\sim$26~\AA\ for the HETDEX spectrum estimate, both greater than the commonly used 20~\AA\ equivalent width minimum as an LAE discriminator \citep{Cowie1998,Gronwall_2007,Adams_2011}.  
In approximately 15\% of the final sample, the imaging counterpart cannot be detected by the Python photometric package (\textit{sep}, \citet{sep_pkg}). However, in all cases, the single emission line is obvious with the assumption that the counterpart is undetected because it is fainter than the imaging limit (here, 26.5 is adopted for HSC-$r$, Figure \ref{fig:summary_stats} panel 5). The classification of the emission line as \lya\, based on the EW analysis, when the continuum is fainter than 25th magnitude is supported (\citet{Leung}; \citetalias{Davis2021, Gebhardt+2021}).

\item \textit{Spatially Resolved Candidates}:
LAEs \textbf{at z$\sim$3} are expected to be a few kpc to maybe 10 kpc in radius \citep{Ribeiro_2016} and should be generally unresolved in the HSC-$r$ imaging (seeing FWHM 0\farcs6-1\farcs0). Any HSC-$r$ detected galaxies with an \textit{sep} elliptical aperture effective radius or ELiXer circular aperture radius \citepalias{Davis2021} $\geq$2\farcs0 ($\gtrsim$ 15 kpc for $z\sim3$) are rejected as they could be misclassified \textbf{an AGN or foreground galaxy} or include \textbf{a slightly offset} line of sight interloper, extending the projected profile. Here the effective radius is defined as:

\begin{equation} \label{eq:Re}
    \frac{1}{2} \sqrt{a b},
\end{equation}
where $a$ is the semi-major axis and $b$ is the semi-minor axis of the aperture ellipse. \textbf{It is also possible that the candidate galaxy is actually two or more interacting LAEs (this is also true even when the candidate is unresolved), but without the higher spatial resolution of $HST$ imaging it is difficult to impossible to distinguish these cases. In regions of the HETDEX survey with multi-band $HST$ coverage this becomes possible using SED comparisons as in \citep{pahl_2021}, but given the large HETDEX fiber size \citep{Gebhardt+2021}, it would challenging to separate the individual LAE spectra. These could be common and are certainly interesting cases for reionization as interactions can lead to enhanced star formation \citep[][and others]{keel_1985,Lawrence_1989,jogee_2009}.} 

\item \textit{Neighbor Flux Contamination}: \label{sec:neighbor_flux}
Regions around known, nearby large galaxies (such as M101 in a region defined by RA (J2000) $\in$ [210.2$^{\circ}$,211.3$^{\circ}$] and Dec (J2000) $\in$ [54.1$^{\circ}$,54.8$^{\circ}$]) are completely excluded from consideration to avoid the risk of contamination from flux contributed by the outer regions of foreground galaxies. Additionally, no candidate is permitted to have any detected object (using the \textit{sep} Python package) extending to within 2$\arcs$ (resolved or unresolved) of the HETDEX detection position. This reduces the chances of emission contamination and source confusion, with the astrometric (centroid) accuracy $\sim$ 0$\farcs$5 and the FWHM seeing generally better than 2$\arcs$ \citepalias{Gebhardt+2021}.

For slightly larger separations, the PSF weighted flux of all imaging found sources within 9 square arcsecs of the candidate LAE are summed and the candidate LAE is removed from the sample if the potential flux contamination exceeds a threshold. As a note pertaining to the execution sequence, this check is performed after the initial query conditions described above and prior to the manual inspections described in the next subsection. The flux for each neighboring source (and the LAE candidate) is computed with simple aperture photometry (again, in HSC-$r$) and convolved with the HETDEX PSF for the observation of the candidate LAE. For simplicity, each source is assumed to be unresolved (noting that larger, extended sources within this area would automatically cause the LAE candidate to be rejected) with a flat spectrum over the band pass. The (moffat) modeled PSF weighted fraction \citepalias{Gebhardt+2021} of flux in the overlap with the LAE candidate and each other source is summed over and the candidate is rejected if that sum \textbf{(over all on-sky neighbors within the 9 square arcsecs)} is larger than 25\% of the candidate LAE's $r$ flux. \textbf{The choice of the 25\% threshold is somewhat rough and is intended to be similar to the average uncertainty in the aperture magnitudes that provide the flux estimates.} For the final sample of $\lyccount$ galaxies, the biweight midvariance location of the fractional PSF weighted flux overlap is 0.01 $\pm$ 0.015, \textbf{far below the 25\% limit}, and that potential flux contribution is ignored. More sophisticated modeling with a flux correction may be used for future samples to allow some relaxation of this criteria.

\end{enumerate}

These initial selection criteria and automated conditions reduce the set of detections from over 1.5 million to approximately 1000 for visual inspection.\\

\subsection{Visual Inspection} \label{subsec:vis_inspection}

The final step in producing the dataset for this work is the manual inspection of the remaining detections that pass the previous selection conditions. Despite best efforts to impose rigorous, repeatable conditions, this is not free from observer subjectivity. Future data selections with improved automation (in development at the time of this writing) will seek to minimize this subjectivity. 

The visual inspection examines the following conditions and excises detections that meet any of the following:
\begin{enumerate}
\item \textit{Poor Imaging}: If the photometric imaging within the few arcseconds around a candidate LAE is unclear, corrupt, incomplete, or shallow (magnitude limit < 25), it cannot be reliably examined for the potential foreground interlopers. Any such detections with poor imaging are removed.

\item \textit{Data Issues}: The 2D HETDEX spectra are examined for cosmic ray strikes, hot pixels, bad columns, or other data corruption in the Lyman continuum spectral region in the nearest four fibers (in order of increasing distance from the detection center position) and in the stacked sum of all contributing fibers. Any issues in these cutouts results in the candidate's exclusion. Similar data issues in the Ly$\alpha$ region do not warrant removal unless they cast doubt on the actual detection or classification as Ly$\alpha$.

\item \textit{Bad Neighbors}: \label{sec:bad_neighbor}
As an extension of the \textit{Neighbor Flux Contamination} condition in the previous subsection, the photometric imaging cutouts for each remaining LAE candidate are examined for the presence of neighboring sources that could be contributing unwanted flux, including any suggestion of very faint, low-surface brightness interlopers that could be missed by the automated source extraction. Larger area (30"$\times$30") zoom-outs of the images are also examined for any potential contaminants (generally, stars or small, but spatially resolved nearby galaxies) that could be sources of scattered light/flux contamination. Again, any candidates with suspect neighbors are excluded.

\item \textit{Unidentified Emission Lines}: The presence of any unidentified emission lines --- that is, any visually apparent or ELiXer Gaussian fitted \citepalias{Davis2021} emission lines in the spectra that are inconsistent with the classification as an LAE at the assumed redshift or suggestive of a second, blended spectrum from a lower-$z$ line of sight interloper --- in the full 1D spectrum suggests some contamination (blending) of spectra. While the ELiXer application performs automated line searching and fitting, this manual check is an additional safety against any convincing emission lines that ELiXer fails to identify. Candidate LAEs with such spectra are rejected. 

\end{enumerate}

\begin{figure*}[ht]
    \centering
    \includegraphics[width=1.0\textwidth]{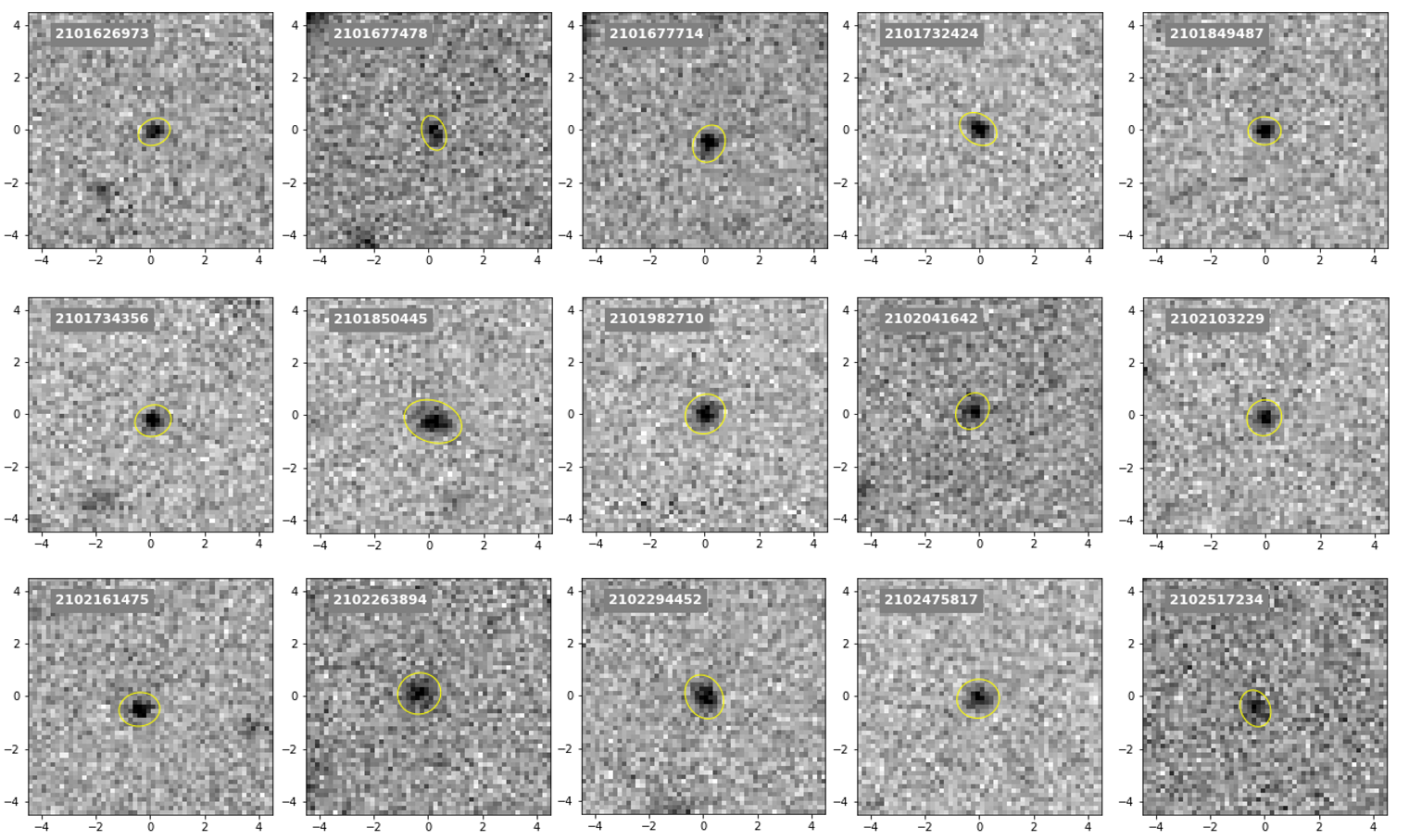}
    \caption{Examples of HSC-$r$ imaging of LAE candidates. The HETDEX identifiers are printed in the upper left and the candidate LAEs are highlighted with gold ellipses defined by the \textit{sep} Python package \citep{sep_pkg}. The axis scale is in arcseconds. The basic data for these candidate LAEs is presented in Table \ref{tab:Table of LAE Galaxies}.  These galaxies are generally unresolved and appear free from foreground interlopers.}
    \label{fig:sample_cutouts_lae}
\end{figure*}

Following this examination, \lyccount\ detections remain to comprise the final dataset for this investigation (again, see Figure \ref{fig:summary_stats} for summary statistics). Figure \ref{fig:sample_cutouts_lae} presents a sub-selection of 15 candidate LAEs as they appear in the HSC-$r$ imaging with Table \ref{tab:Table of LAE Galaxies} listing the basic positional and brightness data for those galaxies (with the full listing made available in a machine readable format). The LAE candidates all appear compact and point source-like with no apparent foreground interlopers. \\

\begin{deluxetable*}{c c c c c c c} [ht]
\tablecaption{Candidate LAE galaxies for this work.  \label{tab:Table of LAE Galaxies}}
\tablewidth{0pt}
\tablehead{
\colhead{Detection ID} & \colhead{Detection Name} & \colhead{RA (degree)} & \colhead{Dec (degree)} & \colhead{$z$} & \colhead{$r$-mag} & \colhead{$r$-err}}
\startdata
2101626973 & HETDEX J131833.31+505207.1 & 199.63881 & 50.86864 & 3.3807 & 25.02 & 0.141 \\
2101677478 & HETDEX J121602.84+504206.1 & 184.01184 & 50.70169 & 3.1182 & 25.06 & 0.145 \\
2101677714 & HETDEX J121552.27+504126.2 & 183.96780 & 50.69062 & 3.3510 & 24.86 & 0.054 \\
2101732424 & HETDEX J123400.37+504626.3 & 188.50156 & 50.77396 & 2.9984 & 24.93 & 0.089 \\
2101849487 & HETDEX J131652.22+504250.0 & 199.21758 & 50.71389 & 3.0323 & 25.53 & 0.052 \\
2101734356 & HETDEX J130453.80+504930.4 & 196.22417 & 50.82512 & 3.2923 & 24.50 & 0.105 \\
2101850445 & HETDEX J131807.08+503938.9 & 199.52948 & 50.66080 & 3.0270 & 24.28 & 0.072 \\
2101982710 & HETDEX J124748.14+562033.2 & 191.95058 & 56.34256 & 3.0517 & 23.93 & 0.070 \\
2102041642 & HETDEX J135500.52+553033.5 & 208.75215 & 55.50930 & 3.1015 & 24.78 & 0.121 \\
2102103229 & HETDEX J142151.25+490031.5 & 215.46356 & 49.00875 & 3.4941 & 25.11 & 0.099 \\
2102161475 & HETDEX J135234.31+504156.7 & 208.14297 & 50.69908 & 3.2663 & 25.02 & 0.085 \\
2102263894 & HETDEX J145659.60+500906.7 & 224.24835 & 50.15186 & 3.2859 & 24.81 & 0.094 \\
2102294452 & HETDEX J130448.56+552515.2 & 196.20233 & 55.42089 & 3.0046 & 23.94 & 0.069 \\
2102475817 & HETDEX J131213.18+542651.4 & 198.05492 & 54.44762 & 3.0704 & 24.37 & 0.094 \\
2102517234 & HETDEX J130019.57+540833.0 & 195.08153 & 54.14250 & 3.2553 & 24.34 & 0.156 \\
... & ... & ... & ... & ... & ... & ... \\
\enddata
\begin{center}
    \small  Basic data for the objects in Figure \ref{fig:sample_cutouts_lae}. The listing for all 214 candidate LAEs for this work is made available in a machine readable format.
\end{center}
\end{deluxetable*}

\section {Analysis} \label{sec:analysis}

The primary objective of this work is to determine the feasibility of detecting Lyman continuum photons in HETDEX data and, as such, the analysis at this stage focuses only on detection with limited consideration given to characterization. Future work will enhance the analysis, correcting for IGM attenuation that are here expressly ignored.

The main input data for this work is the PSF weighted, co-added spectra from the accepted detections selected as described in Section \ref{sec:selection}. Additionally, for each set of exposures, the "empty" or "sky" fibers are stacked and their average residuals are subtracted from the galaxy stacks in a supplemental cleaning step. The HETDEX data reduction pipeline handles all the cleaning, rectification, throughput adjustment, differential atmospheric refraction (DAR) correction, and other steps necessary to combine the individual fibers over multiple exposures into a single 1D spectrum per detection \citepalias{Gebhardt+2021}.\\

\subsection{"Sky Fiber" Selection and Stacking} \label{sky_fiber_stacking}

\begin{figure}[ht]
    \centering
    \includegraphics[width=0.5\textwidth]{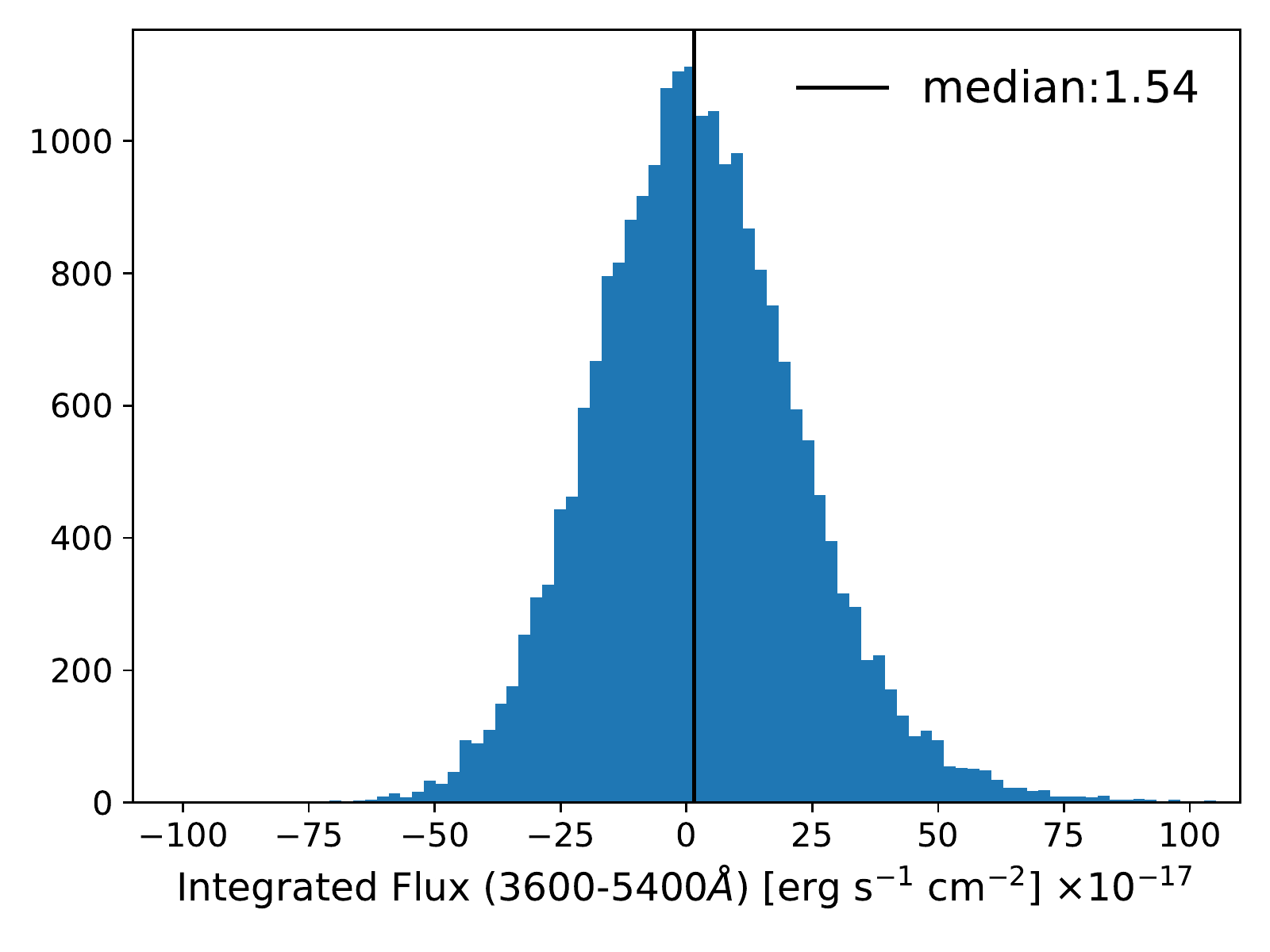}
    \caption{Example histogram of the integrated flux (3600-5400~\AA) in qualifying "sky fibers" (prior to excluding the tails). The fluxes come from observation \#18 on 2020-06-25 with $\sim$88K total fibers and $\sim$23K identified sky fibers. The distribution median is 1.54$\times$10$^{-17}$ \cgs\  with a standard deviation of 21.1 in the same units. See also Section \ref{sky_fiber_stacking}.}
    \label{fig:example_sky_hist}
\end{figure}

"Sky fibers" are effectively empty fibers (i.e. those that do not fall on any apparent astrophysical object) and they play an important role in this work. After HETDEX processing, any residual flux in these fibers is ostensibly due to random noise, including instrumental noise, and imperfect sky measurement and subtraction. While the majority of fibers in the typical HETDEX exposure fall on empty sky \citepalias{Gebhardt+2021}, we are interested in those most free from any measurable light from the PSF wings of nearby objects or otherwise scattered in along the light path as the most representative of the true sky. Functionally, sky fibers, within each three-exposure set, are identified by the following procedure that acts on the post sky-subtracted, rectified spectra:
\begin{enumerate}
    \item To remove fibers with possible measured continuum, any fibers with spectra exhibiting an average flux density (over 500~\angstrom\ wide bins) outside of $\pm$5$\times$10$^{-18}$ [\cgsa] or $\pm$4$\times$10$^{-18}$ [\cgsa] over (almost) the full width (3600-5400~\angstrom) are eliminated. The 3600-5400~\angstrom\ range is selected to avoid the edges of the detector and the prominent sky lines near 3545~\angstrom\ and 5462~\angstrom. The choice of bounding flux densities is an empirical definition and based on the HETDEX flux limits. Since most (50-70\%) fibers fall on empty sky, the values are selected to be small enough to exclude HETDEX-detectable continuum and potential over subtraction of sky \citepalias{Gebhardt+2021} and still include roughly half of all fibers for a typical HETDEX exposure set. Here, by "typical" we mean those exposure sets without large errors (instrument failures) or significant foreground objects (bright stars or relatively nearby galaxies, like M101) that covers a large fraction of the fibers.  
    \item Fibers with potential emission lines are also removed. For expediency, rather than attempt to fit $\sim$1000 Gaussian profiles per single fiber spectrum (with $\sim$100,000 spectra per exposure set), here a \textit{possible} emission line is defined in a sliding window of 3 wavelength bins, totaling 6~\AA\ wide, with the integrated flux over the central wavelength bin exceeding 5$\times$10$^{-17}$ [\cgs] and the two immediately adjacent wavelength bins exceeding 4$\times$10$^{-17}$ [\cgs] each. These emission line flux thresholds are selected based on a typical HETDEX flux limit under good conditions (noting that the actual flux limit is a function of seeing FWHM, throughput, wavelength, CCD response, fiber, etc) \citepalias{Gebhardt+2021}.
    \item The remaining fibers are sorted by their integrated flux over 3600-5400~\angstrom\ and the 1/3 of the fibers in the sum of the two tails of this approximately normal distribution are eliminated \textbf{(i.e. keeping the central 2/3 or roughly 1-$\sigma$). This additional down-selection is simply an extra precaution against the influence of outliers on this residual.} Figure \ref{fig:example_sky_hist} shows a representative example of the distribution of sky fiber flux (before the tails are removed). Here we display the histogram, peaking very close to zero, of the summed fluxes in the 23K identified sky fibers in exposure set \#18 taken on June 25, 2020.
    \item The remaining fibers (roughly 30\% of all fibers or about $\sim$25k fibers per exposure set) are considered effectively empty.
\end{enumerate}

\begin{figure*}[ht]
    \centering
    \includegraphics[width=1.0\textwidth]{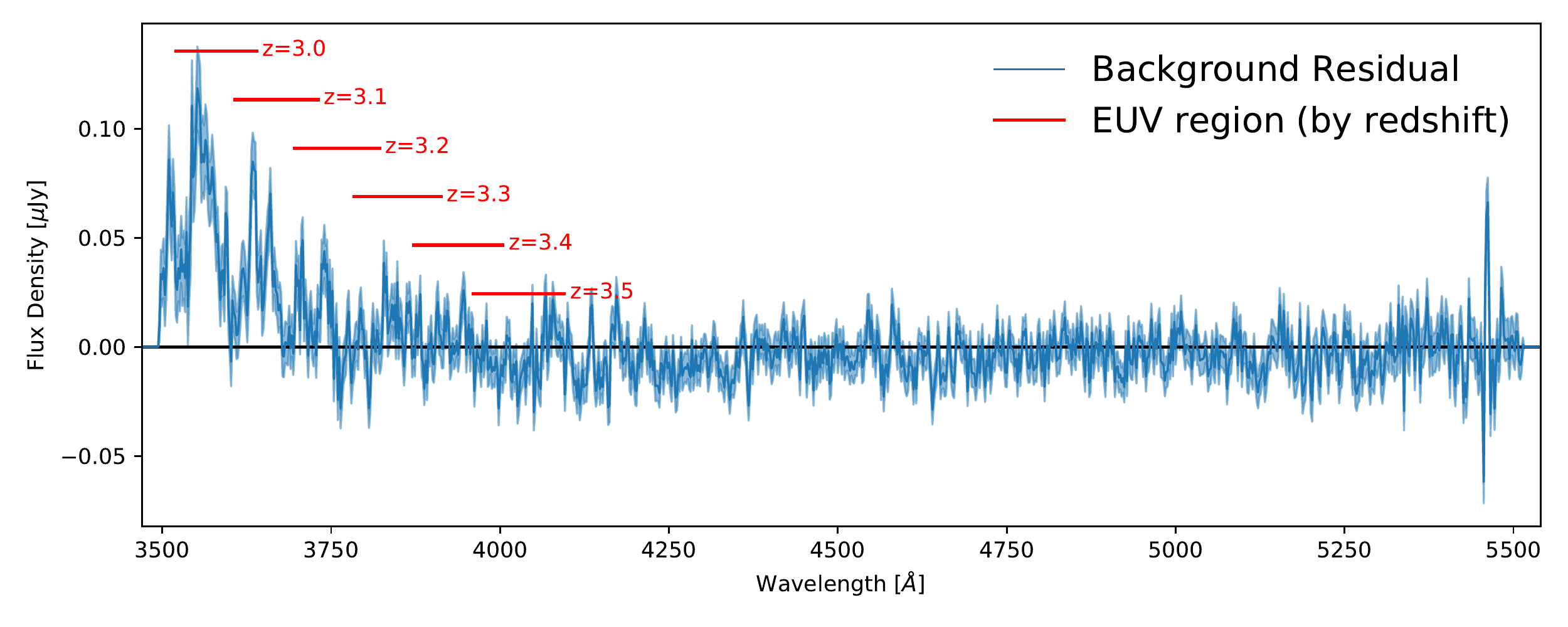}
    \caption{Example biweight of the background residual spectrum (observation \#18 from 2020-06-25). The red horizontal lines indicate the observed frame spectral region that corresponds to 880-910~\angstrom\ at the labeled redshift. The separation along the flux density axis is artificial. There is a clear increase in the background residual flux in the blue end, primarily due to instrument effects \citepalias{Gebhardt+2021}.
    The background residual for each specific observation is subtracted from the corresponding galaxy spectra prior to stacking. For a discussion of the selection of "sky fibers" and the creation of background residual spectra, see Section \ref{sky_fiber_stacking}.}
    \label{fig:example_sky_spec}
\end{figure*}

If the background subtraction was perfect, the distribution would center on zero (it is very close, Figure \ref{fig:example_sky_hist}) and no wavelength bins or ranges of bins should vary from zero with anything but random noise. To be clear, we use the term "background" to encompass both the true sky background and instrumental noise.  Figure \ref{fig:example_sky_spec} reveals, however, there is some small average residual, particularly to the blue end of the spectrum that must be corrected. This figure (the stacked background residual spectrum from exposure set \#18 taken on 2020-06-25) is typical of HETDEX observations in the fraction of sky fibers and in the residual spectrum. Even though there are larger residuals in the blue, the peak near 3550~\AA\ and 3$\times$10$^{-19}$ [\cgsa] is still far below the stacked, residual subtracted, galaxy flux in the Lyman continuum region presented later in this paper.

Stacking of the spectra in the defined sky fibers (again, within each data sample included exposure set) is performed on a per-wavelength basis, with the wavelength bins aligned on their "native" observed frame (2~\angstrom) boundaries, using a biweight scheme, \textbf{not a sum or average}. The biweight measure of central location \citep{Beers_1990} is similar to a median and is selected for its stability and robustness against outlier influence, \textbf{reducing or eliminating any bias toward the relatively few, brighter sources}. The biweight implementation used in this work is slightly modified with an additional weight applied as the inverse uncertainty in the flux measurements. Hereafter, we refer to this modification of the standard biweight location ($\zeta_\mathrm{biloc}$) simply as the "weighted biweight".

\begin{align}
    \zeta_\mathrm{biloc} &= M + \frac{\sum_{v_i>0}(x_i-M)v_i^2}{\sum_{v_i>0}v_i^2},
\end{align}
where $M$ is an initial guess (typically the sample median) and the weights, $v_i$, are defined as 
\begin{equation}
    v_i = 1-u_i^2
\end{equation}
and 
\begin{equation}
    u_i = \frac{x_i - M}{c\times \mathrm{median\left\{|x_i-M|\right\}}}
\end{equation}
where $x_i$ are the data and $c$ a tuning constant (commonly $6$, see \citet{Beers_1990}).

The biweight location "shifts" the median by the average difference of points to the median $(x_i-M)$. This average excludes outliers and assigns higher weights to points close to the median than points that are farther away.
As this shift is a weighted average, we can include the additional inverse uncertainty weights $w_i$ by modifying the weights $v_i^2$:
\begin{equation}
    v_i^2 \rightarrow v_i^2 + \widetilde{w_i}^2  ,
\end{equation}
where
\begin{equation}
    \widetilde{w_i} = \frac{w_i}{2\times \mathrm{median\left\{w_i\right\}}}
\end{equation}
This "normalization" of the additional weights, $w_i$, is necessary to ensure that $v_i$ and $\widetilde{w_i}$ have comparable values and therefore similarly large contributions to the final weights.
The weighted biweight location becomes:
\begin{equation}
    \widetilde{\zeta}_\mathrm{biloc} = M + \frac{\sum_{v_i>0}(x_i-M)(v_i^2+\widetilde{w}_i^2)}{\sum_{v_i>0}(v_i^2+\widetilde{w}_i^2)}
\end{equation}

\subsection{Background Residuals Correction} \label{subsec:background_corr}
Although great pains are taken to remove the background signal from the galaxy spectra in the primary sky subtraction steps, two additional steps are taken to further refine this correction. First, to more consistently subtract the background spectrum from the galaxy spectra, each of the LAE candidate galaxy spectra are re-extracted from the original HETDEX data using the full-field sky subtraction to replace the IFU local sky subtraction used in the original iHDR2 catalog. Here, "full-field" refers to all fibers in all IFUs for an exposure set and "local" refers only to those fibers in the single IFU where the galaxy detection is made. This approach greatly expands the number of fibers used to compute the sky and reduces the potential for over-subtraction that can result from diffuse (or stray) light in a large fraction of the fibers in a single IFU (\citetalias{Gebhardt+2021}; \citet{Zeimann_2021}).

Second, for every galaxy spectrum, the independently selected sky fibers (as previously described) for all IFUs and exposures that contributed to the calculated full-field sky background for that exposure set are collected and averaged (again, using a weighted biweight). The specific PSF model for that exposure set is then applied to the resulting residual background spectrum and subtracted from the galaxy spectrum in the observed frame. Logically, this is equivalent to subtracting this background residual from the individual galaxies fibers before they are PSF weighted and combined. An example of the background residual from one exposure set (\#18 from 2020-06-25) is shown in Figure \ref{fig:example_sky_spec}.

In an additional check, the region around the particularly bright sky line at 3535-3555\AA\ is masked from the galaxy spectra. The resulting galaxy \textbf{weighted biweight} stacks and Lyman continuum region averages (not shown) exhibit no discernible differences, so no extra correction for this sky line (beyond those described above) is applied.\\

\subsection{Galaxy Spectra Stacking}

Generally, aside from the primary \lya emission line, the signal-to-noise ratio of astrophysical photons over other wavelength ranges for our $z\gtrsim 3$ LAEs is too low for detection in individual sources. We therefore employ stacking to boost the signal. Since the observed \lya line is found at different wavelengths (4860 to 5500~\angstrom), the cleaned and corrected galaxy spectra are first redshifted to their own restframes, as determined from the \lya line, so they can be aligned for \textbf{the weighted biweight} stacking. The individual redshifting produces wavelength bins of slightly different widths, from 0.44 to 0.50$~\angstrom$. The largest redshift results in the most narrow bins and is adopted as the grid onto which all other rest-frame spectra are linearly interpolated. The spectra are then aligned on the Ly$\alpha$ bin, for now ignoring systemic Ly$\alpha$ velocity offsets \citep{Verhamme_2018} (discussed later), and the weighted biweight is computed for each wavelength bin. The uncertainty is reported as a standard error on the biweight location, with the square root of the biweight midvariance (or biweight scale) as $\sigma$.  The relatively small uncertainty, roughly 0.05~\AA\ to 0.12~\AA\ in the rest-frame, in the Gaussian fit line-center \citepalias{Gebhardt+2021} is also ignored as it is much smaller than the systemic velocity error and, unlike the velocity error, averages to zero over the sample.  As with the sky fibers, averaging is performed with the modified (weighted) biweight. 

\begin{figure*}[ht]
    \centering
    \includegraphics[width=1.0\textwidth]{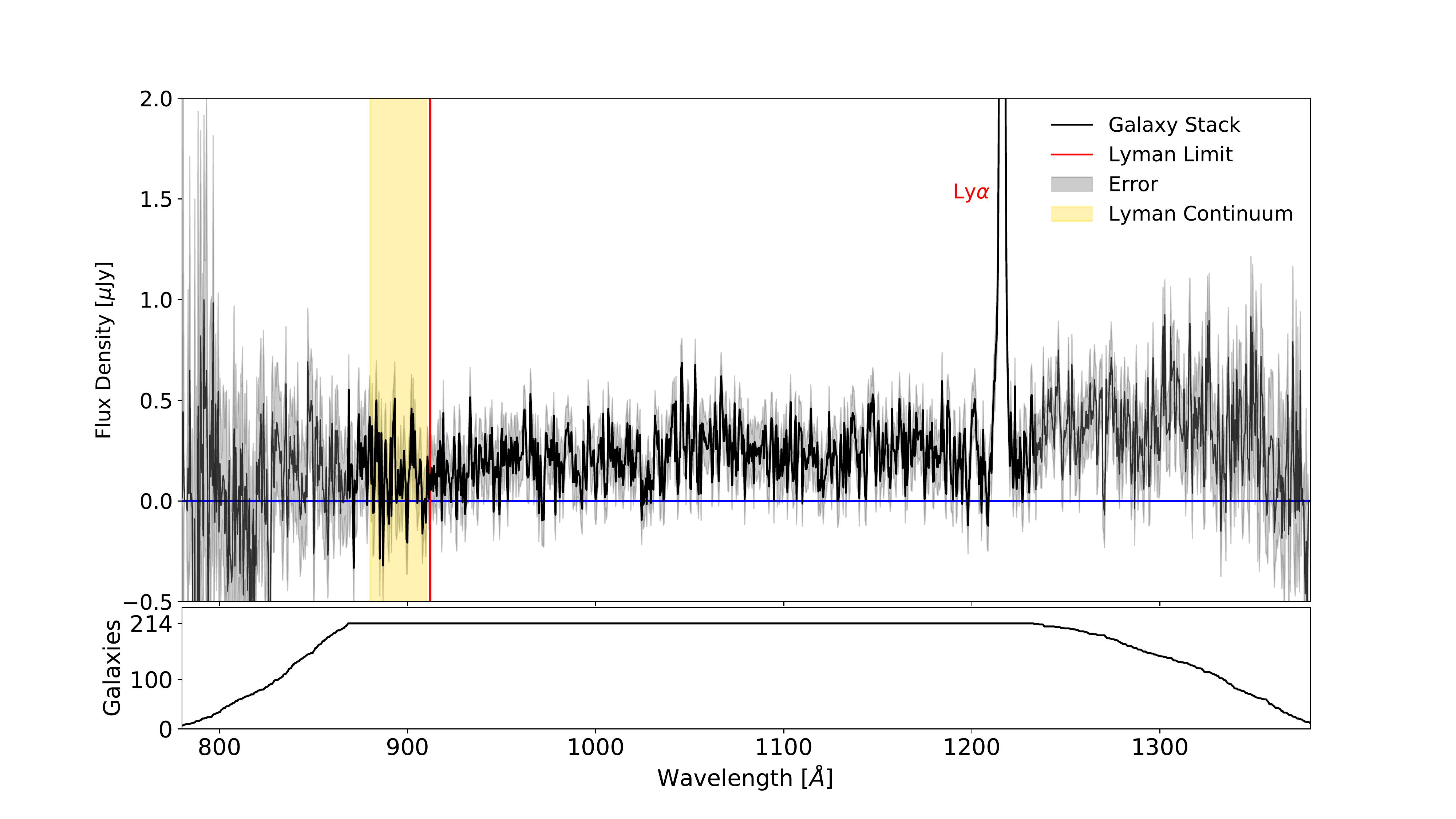}
    \caption{The restframe stack of the galaxy spectra, with correction for background residuals but without correction for IGM attenuation. The lower panel shows the number of galaxies from the sample that contribute to the corresponding wavelength bins in the upper panel. Due to the fixed observed frame wavelength range and the variation in redshift of the galaxies, the wavelength regions blueward of the marked Lyman continuum and redward of Ly$\alpha$ have fewer contributors. }
    \label{fig:flam_stack}
\end{figure*}

Figure \ref{fig:flam_stack} (top panel) presents the result of stacking the full sample of candidate LAEs. Because of their different redshifts but uniform observed frame wavelength range, the wavelength bins in the extreme outer blue and red ends of the \textbf{weighted biweight} stack, well beyond the Lyman continuum and Ly$\alpha$ regions, include fewer data points (lower panel) and are truncated from the next figures. 

Figures \ref{fig:main_plot} and \ref{fig:main_plot_zoom} show the main galaxy spectra stacks for this work, with the wavelength bin data listed in Table \ref{tab:stack_spectra_listing}. To reduce the jitter and boost SNR, the final galaxy \textbf{weighted biweight} stacks are summed over bins 11 elements wide ($\sim$5~\angstrom\ given the uniform restframe bin width of $\sim$0.45~\angstrom). The black line represents the binned stack of the entire sample. Also shown are binned stacks of the brightest 1/3 (in blue) and faintest 2/3 (red) by $M_{\text{UV}}$. 

In all three \textbf{weighted biweight} stacks some continuum on either side of Ly$\alpha$ is clearly present (at least red-ward of the Lyman Limit) and absorption for some of the Lyman Series are apparent. Other common absorption features are also marked, with several prominently represented which will be useful for future stellar population fitting. Of note is what appears to be a P-Cygni profile of \ovi~($\lambda$1032,1038\AA) as a signature of the young stellar populations \citep{Leitherer_1999} in these LAEs.

As is expected simply by the selection, the brightest 1/3 shows increased continuum over the faintest 2/3 until near and blue-ward of the Lyman Limit, where the bright subsample falls off rapidly. This could be consistent with an expected trend to a higher EUV escape with decreasing UV luminosity (and decreasing halo mass) \citep{Nakajima+2018,Finkelstein_2019}, \textbf{though the faint-end limit of this sample is far too bright and the sample size far too small for any conclusions.} At a minimum, the absence of a statistically significant detection of EUV emission in the $M_\text{UV}$ brightest 1/3 vs the clear detection in the fainter subsamples suggests an $M_\text{UV}$ limit where EUV escape declines rapidly. Even with the decreased HETDEX sensitivity to the blue end of the spectra range, the Lyman continuum region in the full sample \textbf{weighted biweight} stack of galaxy spectra from this work is clearly above and inconsistent with zero flux (Figure \ref{fig:main_plot_zoom}), supporting the claim of flux detection in this limited sample.\\

\subsection{Lyman Continuum Averaging}\label{stacking_sequence}

Three averaging methods are applied to the Lyman continuum region with somewhat different detection significances but with consistent results. In the first method, the weighted biweight measure of central location, hereafter "weighted biweight", of the flux density of the Lyman continuum region for each individual (rest-frame) spectrum is computed and the weighted biweight of those individual EUV flux densities is calculated. For this method, the result, 0.10 $\pm$0.036 $\mu$Jy, has the lowest significance, in part due to the larger error from the fewer wavelength bins ($\sim$66 per rest spectrum), but is the most straightforward as the Lyman continuum regions from the individual spectra are taken as whole inputs. It is this result that is reported as the primary finding as it is the most conservative of the three methods. In the second method, all Lyman continuum region wavelength bins from all sample spectra are placed in a single vector ($\sim$14,000 elements) and the weighted biweight is computed. The results are consistent, 0.11 $\pm$0.024 $\mu$Jy, with a larger significance. In the third method, the weighted biweight of the Lyman continuum region of the resulting spectrum from the full stack (described above) is computed. Again, the results are consistent, 0.12 $\pm$0.021 $\mu$Jy with the highest significance of the three methods, but due to the weighted biweight of the stacking, each wavelength bin from individual spectra does not necessarily receive the same weighting, and so a given spectrum may contribute somewhat differently to each $\sim$0.5~\AA\ bin in the final stack. However, this is only a small caveat in the interpretation and does not affect the conclusions, noting that all three methods provide very similar statistics.

This same procedure is repeated for a few sub-selections of the data and the results are presented in Table \ref{tab:EUV_subselect_summary}. As noted in the previous section, the trend to a higher absolute EUV emission with decreasing $M_{\text{UV}}$ persists even with this small sample size and low significance. This result might support models of Reionization that favor an early start with most of the ionizing radiation budget supplied by numerous, fainter star-forming galaxies (see \citet{Finkelstein_2019} for an extensive discussion). Additional data will allow for substantial refinement in both SNR and luminosity bin resolution.\\

\begin{deluxetable*}{l r c c c c c} [ht]
\tablecaption{Summary of Sub-sample Properties with EUV (900\AA) Emission \label{tab:EUV_subselect_summary}}
\tablewidth{0pt}
\tablehead{
{Description} & {Count} & {$\tilde{z}$} & {$\tilde{M}_{\text{UV}}$} &  {Method 1} & {Method 2} & {Method 3} \\
 & & & $\pm\sigma$  & ($\mu$Jy) $\pm$SE &  ($\mu$Jy)  $\pm$SE &   ($\mu$Jy)  $\pm$SE
}
\startdata
Full Sample   & 214 & 3.16 & -20.35 $\pm$ 0.711 & 0.10 $\pm$ 0.036 & 0.11 $\pm$ 0.024 & 0.12 $\pm$ 0.021\\
Brightest 1/3 & 72  & 3.15 & -20.99 $\pm$ 0.273 & 0.01 $\pm$ 0.054 & 0.01 $\pm$ 0.043 & 0.02 $\pm$ 0.037\\
Faintest 2/3  & 142 & 3.17 & -20.00 $\pm$ 0.532 & 0.15 $\pm$ 0.045 & 0.15 $\pm$ 0.029 & 0.18 $\pm$ 0.026\\
Faintest 1/3  & 72  & 3.18 & -19.40$^{+}$ $\pm$ 0.327 & 0.13 $\pm$ 0.066  & 0.12 $\pm$ 0.041  & 0.14 $\pm$ 0.035\\
\enddata
    \begin{center} \small $^{+}$Note: The Faintest 1/3 $\tilde{M}_{\text{UV}}$ may be fainter as 30 of the 72 galaxies are at or below the $r$ imaging flux limit.
    \end{center}
    Summary of basic properties of sample sub-selection (note: the \~{} overscore indicates a median). The redshift distribution of each subsample are effectively the same with almost identical median and scatter over [3.00,3.49]. All reported statistics are the weighted biweight with the error on $M_{\text{UV}}$ as the sub-sample biweight scale. The error on the three EUV escape columns are the standard errors. Method 1 computes the flux density in the Lyman continuum region of each individual candidate LAE spectrum and then performs a weighted biweight over those flux densities. Method 2 concatenates all Lyman continuum wavelength bins across all LAE candidates and performs a weighted biweight on the resulting single vector. Method 3 aligns the rest-frame wavelength bins of all candidate LAE spectra and stacks with a weighted biweight and then computes the flux density in the Lyman continuum region of the resulting stacked spectrum. See Section \ref{stacking_sequence}.
\end{deluxetable*}

\begin{figure*}[ht]
    \centering
    \includegraphics[width=1.0\textwidth]{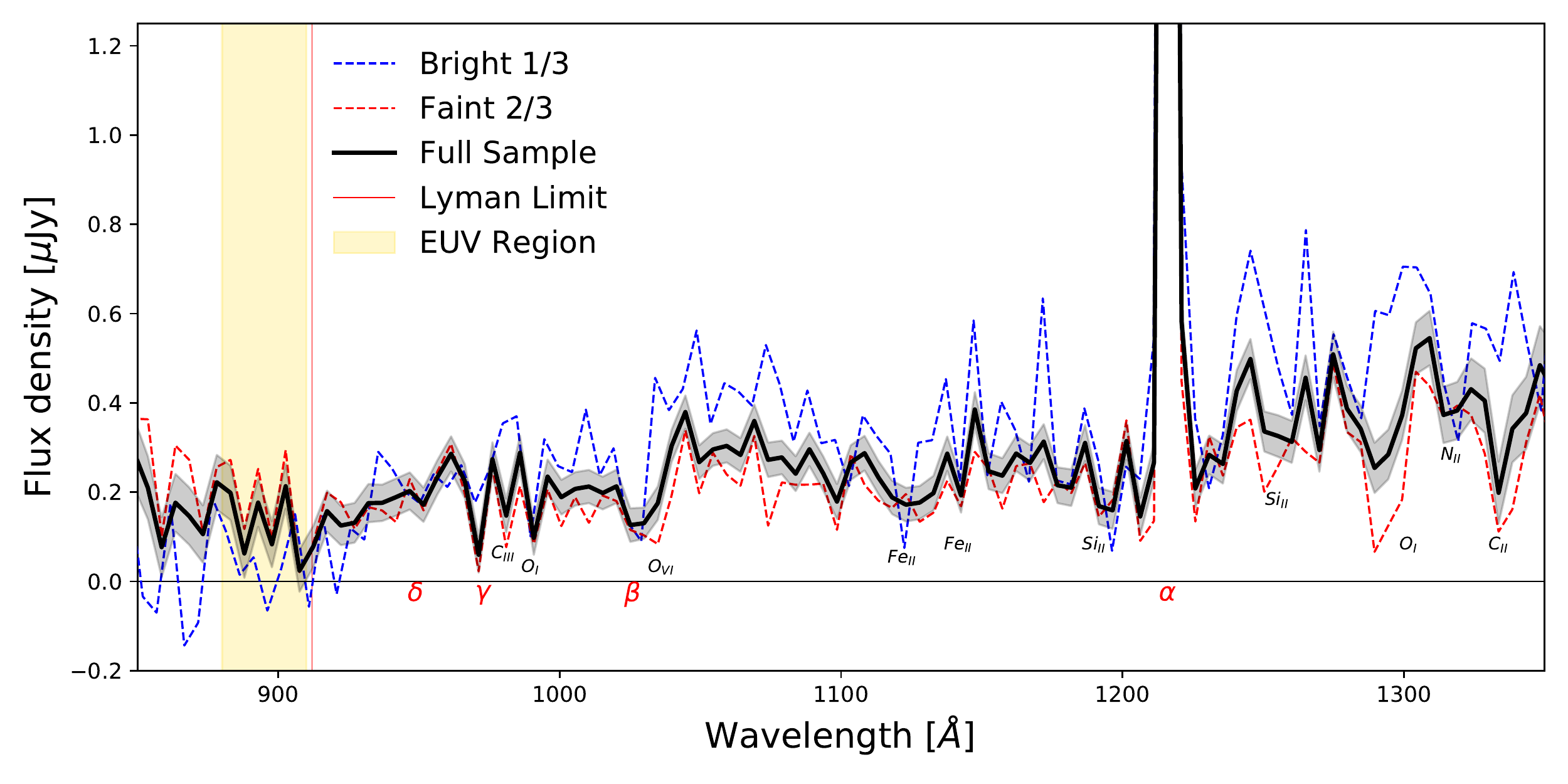}
    \caption{Flux density (presented in $\mu$Jy here) \textbf{weighted biweight} stack of all \lyccount\ galaxy spectra (without a systemic \lya\ velocity correction) binned to $\sim$5~\angstrom. Also shown are the brightest 1/3 (median $M_{\text{UV}}$ -21.0) and faintest 2/3 (median $M_{\text{UV}}$ -20.0) along with the locations of the Lyman Series (in red text) and some common absorption features (in black text). For readability, the Bright 1/3 and Faint 2/3 stacks are shown without their errors. See Figure \ref{fig:main_plot_zoom} for an expansion of the Lyman continuum region.}
    \label{fig:main_plot}
\end{figure*}

\begin{figure}[ht]
    \centering
    \includegraphics[width=0.5\textwidth]{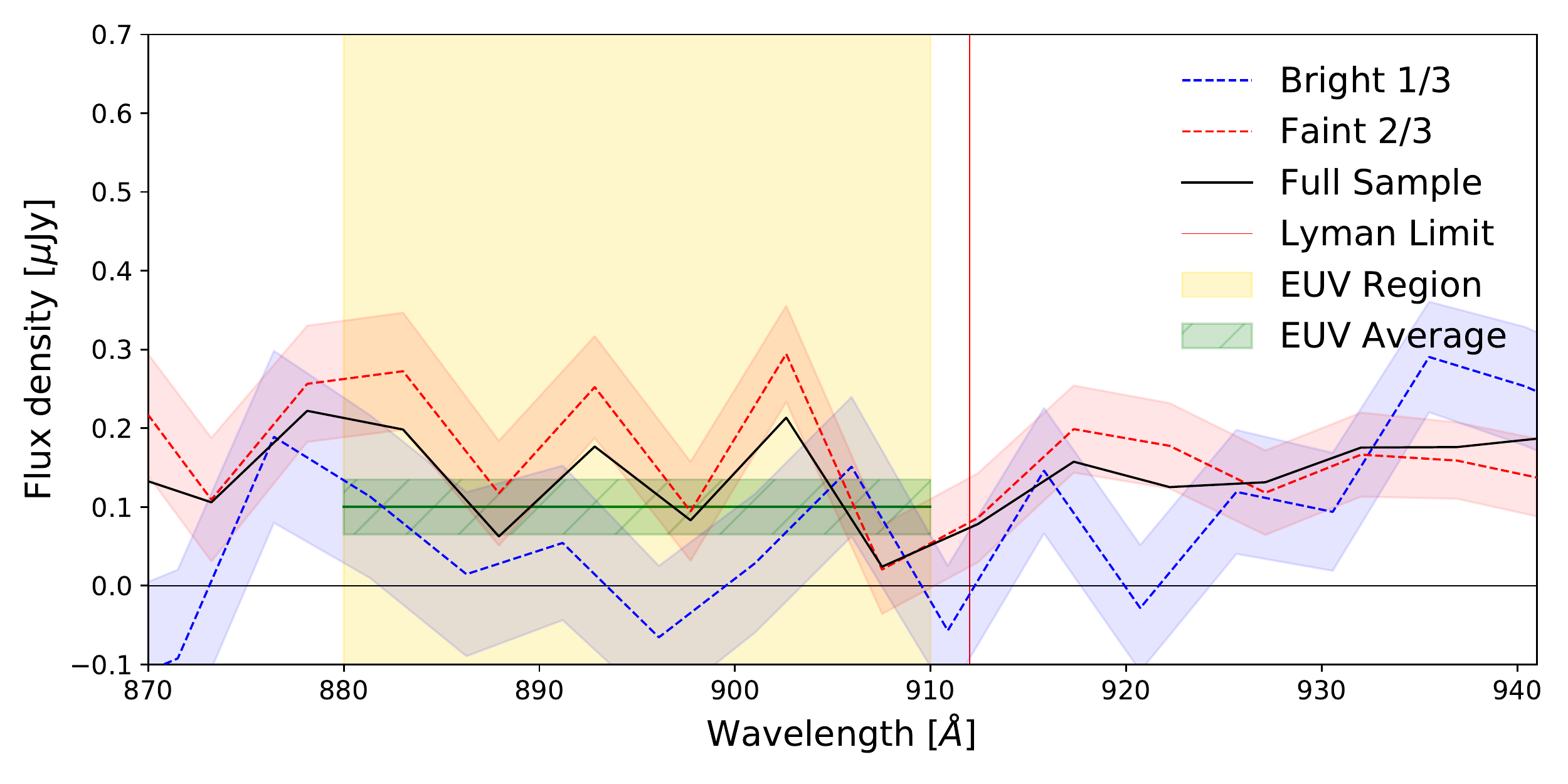}
    \caption{Expanded view of the Lyman continuum region (Figure \ref{fig:main_plot}). For readability, the Full Sample \textbf{weighted biweight} stack is shown without its error. The flux (presented in $\mu$Jy here) in the Lyman continuum region for the full sample is well above and inconsistent with the zero flux line. The full sample weighted biweight of individual EUV flux densities, Method 1, (shown as the green horizontal line) is \lycujy\ (see Section \ref{stacking_sequence}). The separation between the Bright 1/3 stack and the Faint 2/3 stack suggests a \textbf{possible} trend of increasing Lyman continuum emission with decreasing host UV luminosity \textbf{over the limited range of M$_{UV}$ of this work.}}
    \label{fig:main_plot_zoom}
\end{figure}

\begin{deluxetable*}{c c c | c c c | c c c | c c c} [ht]
\tablecaption{Sub-sample Stacked Spectra \label{tab:stack_spectra_listing}}
\tablewidth{0pt}
\tablehead{
 & {Full Sample} & &  & {Brightest 1/3} &  &  & {Faintest 2/3} &  &  & {Faintest 1/3} &  \\
{$\lambda$} & {Flux} & {Flux Err} & {$\lambda$} & {Flux} & {Flux Err} & {$\lambda$} & {Flux} & {Flux Err} & {$\lambda$} & {Flux} & {Flux Err} 
}
\startdata
... & ... & ... & ... & ... & ... & ... & ... & ... & ... & ... & ... \\
878.13 & 4.2299 & 1.1694 & 876.43 & 3.6287 & 2.0956 & 878.13 & 4.8827 & 1.4072 & 877.32 & 3.6657 & 1.9114 \\ 
883.03 & 3.7358 & 1.1480 & 881.35 & 2.1448 & 1.9637 & 883.03 & 5.1311 & 1.4022 & 882.23 & 3.2228 & 1.9740 \\ 
887.93 & 1.1649 & 1.0221 & 886.27 & 0.2733 & 1.9557 & 887.93 & 2.1873 & 1.2408 & 887.14 & 0.8911 & 1.7336 \\ 
892.83 & 3.2563 & 0.9949 & 891.19 & 1.0077 & 1.8225 & 892.83 & 4.6475 & 1.1933 & 892.05 & 1.6511 & 1.7201 \\ 
897.73 & 1.5130 & 0.9396 & 896.11 & -1.2052 & 1.6673 & 897.73 & 1.7214 & 1.1473 & 896.95 & 1.7507 & 1.5381 \\ 
902.62 & 3.8480 & 0.9080 & 901.04 & 0.5228 & 1.6001 & 902.62 & 5.3071 & 1.1018 & 901.86 & 4.7167 & 1.5161 \\ 
907.52 & 0.4292 & 0.8444 & 905.96 & 2.7152 & 1.5920 & 907.52 & 0.3651 & 1.0064 & 906.77 & 2.3884 & 1.4004 \\ 
912.42 & 1.3796 & 0.8228 & 910.88 & -1.0136 & 1.4604 & 912.42 & 1.5202 & 0.9989 & 911.68 & 1.1720 & 1.4586 \\ 
... & ... & ... & ... & ... & ... & ... & ... & ... & ... & ... & ... \\
\enddata
\begin{center}
Spectra of the galaxy stacks from Table \ref{tab:EUV_subselect_summary} and Figures \ref{fig:main_plot} and \ref{fig:main_plot_zoom}.
\end{center}
 The center wavelength of each bin ($\lambda$) is reported in \AA\ and the Flux and Flux Err are in erg s$^{-1}$ cm$^{-2}$ $\times$10$^{-17}$ integrated over the $\sim$5~\AA\ wide wavelength bins. Shown here are the data in the Lyman continuum region of each \textbf{weighted biweight} stacked spectrum. The complete listing is made available in a machine readable format. Note: for easier comparison to the reported EUV emission flux densities, Figures \ref{fig:main_plot} and \ref{fig:main_plot_zoom} are plotted in $\mu$Jy instead of the CGS units in this table. 
\end{deluxetable*}

\subsection{Lyman Alpha Line Dispersion and Systemic Velocity Offset}  \label{sec:velocity_offset}

The complex resonant scattering of \lya photons (\citet{Field}, \citet{Moses}, and others) means that the emission line typically separates into a blue peak and a red peak, with the blue peak often highly suppressed, resulting in a (sometimes large) line dispersion from the many scatterings and a systemic velocity offset to the galaxy when fitting only the red peak. This can create a problem when aligning the spectra for stacking and certainly for any precise characterization of the \lya line. These restframe velocity offsets can be in the hundreds of km s$^{-1}$ \citep{Shapley_2003,Berry_2012,Verhamme_2018} and are generally correlated with the \lya line FWHM. The low spectral resolution of VIRUS (R$\sim$800) amounts to an emission line minimum FWHM resolution of $\sim$375 km s$^{-1}$ (\citetalias[][]{Gebhardt+2021}; \citet{Zeimann_2021}) for our \lya lines. This is sufficient for line resolution, except for the most narrow emission lines (39 of 214 \lya\ lines are below 375 km s$^{-1}$ FWHM). Velocity offsets to the fitted line center are measurable with greater precision ($\sim$30 km s$^{-1}$, or $\Delta z \sim$ 0.0005 \citepalias{Gebhardt+2021}).

With the moderate range of luminosities and line widths of the LAEs in this work and the more modest objective of simply detecting Lyman continuum emission, combined with the relatively broad spectral region over which the Lyman continuum was averaged, these complications are largely ignored. With that caveat, an approximate correction for the systemic velocity offset of the red peak of \lya\ (V$^{\rm{red}}_{\rm{peak}}$) from its host galaxy is explored with equation 2 from \citet{Verhamme_2018}: 

\begin{equation}
    \rm{V}^{red}_{peak} = 0.9(\pm0.14) \times \rm{FWHM_{Ly\alpha}} - 34(\pm60)\ \rm{km\ s^{-1}}
\end{equation} 

Applying this relation to the individual spectra prior to stacking, and using the 375 km s$^{-1}$ minimum FWHM resolution for those lines below that limit, produces a $\lesssim$ 0.1\% overall change in individual redshifts, around 200 km s$^{-1}$ averaged over the entire sample (well above the HETDEX line center precision), and results in no significant difference in the Lyman continuum flux density using any of the averaging methods. Methods 1,2, and 3 exhibit changes of +2\%, +5\%, and +10\% respectively. These corrections are all far less than the uncertainty in the weighted biweight of the Lyman continuum emission and have no meaningful impact on the detection at this level. With the current small sample size and moderate scatter in \lya\ line FWHM, the spectral stack is constructed from a somewhat inhomogeneous sample, and a singular correction applied post-stacking would not be appropriate. However, with expanded future datasets allowing for stacks of larger numbers of more uniform luminosities, \lya\ line widths, etc, the velocity correction, in particular, could become more important in resolving ensemble spectral features and will be revisited.

\begin{figure}[ht]
    \centering
    \includegraphics[width=0.5\textwidth]{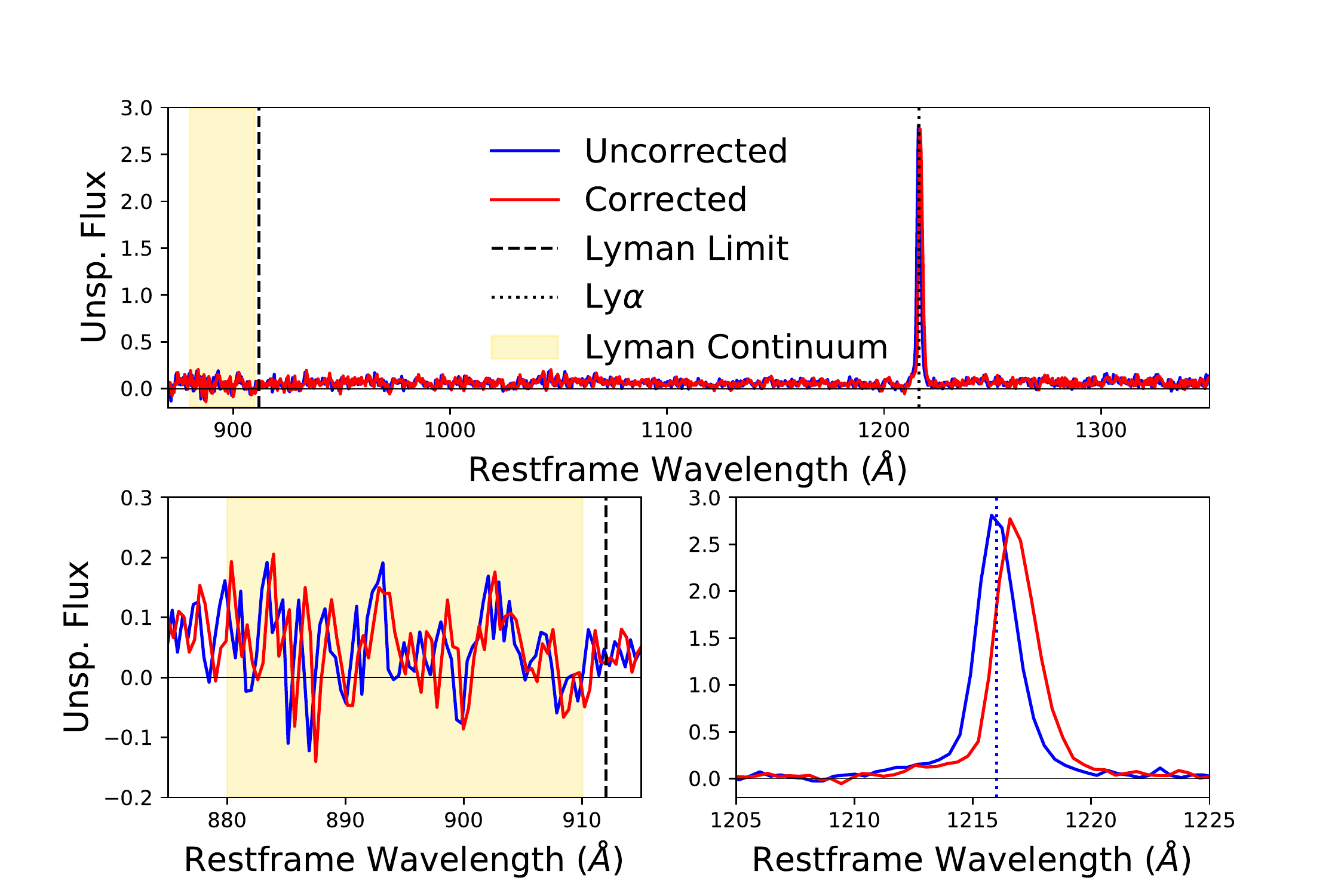}
    \caption{Galaxy spectra \textbf{weighted biweight} stack with (red) and without (blue) approximate velocity corrections \citep{Verhamme_2018}. The averaged velocity offset of the stack was $\sim$200 km s$^{-1}$ (to the red), well above the HETDEX line center sensitivity ($\pm$30 km s$^{-1}$), and makes no measurable difference in the computed Lyman continuum flux density.}
    \label{fig:verhamme_correction}
\end{figure}

Figure \ref{fig:verhamme_correction} presents an overview of the Ly$\alpha$ velocity correction. The top panel shows essentially the full width view of the full-sample LAE \textbf{weighted biweight} stack with the corrected spectrum (red) over-plotted with the uncorrected spectrum (blue). The lower two panels highlight the Lyman continuum region (left) and the Ly$\alpha$ line (right), respectively. The shift in the line center is clearly visible in the lower-right panel, but is still relatively small and makes no significant difference over the Lyman continuum region (lower-left panel).\\

\subsection{Error Systematics}
Systematics in error analysis are difficult to quantify and can have a particularly large impact with smaller datasets. In an effort to report the most accurate detection significance, the following procedure (comparing subset standard errors on the mean to the standard deviations of the means of the subsets) is used to estimate a correction for the combined systematics (statistical, measurement, and instrument). 
\begin{enumerate}
  \item Randomly subdivide the sample into (\textit{m}) non-overlapping, non-repeating proper sub-samples of size (\textit{n}) (where \textit{n} is $\leq$ integer of the total sample size divided by \textit{m}).
  \item Create a single vector of fluxes over the target wavelength range (i.e. 880-910~\AA\ for the Lyman continuum region) across all (\textit{n}) spectra for each of the (\textit{m}) subsets and take the standard error of those fluxes ($\sigma_{n}/\sqrt{n}$). Here \textit{n} is actually the subset size $\times$ the number of wavelength bins (which varies slightly from 65 depending on the subset maximum redshift), but for compactness of notation is represented by the number of galaxies in the subset.
  \item Compute the mean of each of the (\textit{m}) subsample flux vectors and take the standard deviation of those means ($\sigma_{m}$).
  \item Repeat steps (1)-(3) 10000 times with different random subset draws.
  \item Compute the mean of the $\sigma_{m}$ and $\sigma_{n}/\sqrt{n}$ values over those 10000 draws and take the ratio of those means ($\overline{\sigma}_{m}$ / ($\overline{\sigma}_{n}/\sqrt{n}$)).
  \item Repeat (1)-(5) for each combination of (\textit{m}, \textit{n}) (down to a minimum \textit{m} of 2).
  \item Fit a line through the ratios (indexed by \textit{n}) and extrapolate to the full dataset to find the systematic correction factor (to divide into the full dataset standard error).
\end{enumerate}
For this dataset, the ratio ($\overline{\sigma}_{m}$ / ($\overline{\sigma}_{n}/\sqrt{n}$)) fit reaches 1:1 near (\textit{n}=82 (for \textit{m}=2)) indicating that the sample is already sufficiently large to not need an additional systematics correction applied to the standard error.\\

\subsection{Spectral Contamination}
This work assumes that there is no contamination of the measured flux resulting from mis-classification of the emission line, PSF scattered light from sky-adjacent astrophysical objects, or extra contributions from line of sight, low surface brightness interlopers. While multiple steps were taken to exclude these potential contaminants from the dataset (see Section \ref{sec:selection}), no additional corrections, beyond the background residual subtraction (Section \ref{sky_fiber_stacking} ), are applied to the individual or stacked spectra. However, the galaxy stack is examined for indications of emission lines that would suggest the presence of mis-classified \OII ($\lambda$3727) or \OIII ($\lambda$5007), \textbf{for example}, within the dataset. The galaxy spectra are aligned and stacked \textbf{assuming the restframe wavelength of each potential mis-classified emission line}, without a correction for any velocity offset (as would be appropriate for these non-resonant line contaminants), and checked for emission lines supporting the assumption that either of these contaminants are present. No such emission is identified. Additional contaminants, \MgII ($\lambda$2795+$\lambda$2802) and \CIII ($\lambda$1909), are also specifically considered. The \CIII line does not appear as a redshifted single line within the observed spectral range considered by this work (4860 - 5540~\AA) and would be expected to be accompanied by \CIV ($\lambda$1549), which is not found. The \MgII doublet (at the edge of resolution as separate lines by VIRUS \citepalias{Gebhardt+2021}) likewise does not appear by itself for any part of the wavelength range of interest and should be accompanied by \CII ($\lambda$2326) over the entire range and by \CIII after \MgII passes 5088~\AA. The \MgII line is often broad and frequently (but not always) associated with AGN and the maximum emission line FWHM cut of 10~\AA\ (see Section \ref{sec:selection}) and absence of continuum in individual galaxy spectra made this an unlikely contaminant. \MgII can also be narrow and appear similar to \lya, however, for the narrow cases, no evidence of either \CII or \CIII are found in the sample, nor do any of the line shapes suggest a doublet peak, favoring the bluer (2795~\AA) peak as is expected for \MgII \citep{Chisholm2020}.\\

\section{Discussion} \label{sec:discussion}

As a qualitative reference and for some context, we select the Keck Lyman continuum Spectroscopic Survey (KLCS) \citep{Steidel_2018} as a representative of prior studies and perform a limited comparison. Additional $z\sim 3$ surveys \citep[including][among others]{Vanzella_2015,Shapley_2016,bian2017,Rivera_Thorsen_2019} will be discussed in the future, examining sample differences and trends, and KLCS is presented here primarily as a sanity check on the selection and stacking methods presented in this work.

\subsection{Comparison to the Keck Lyman Continuum Spectroscopic Survey}\label{vs_klcs}

Since the continuum selected KLCS sample median is $\sim$2$\times$ brighter (in \textbf{$r$ and} $M_{\text{UV}}$) than that of the galaxies in our sample, the 52 brightest galaxies from this work are chosen to stack as their median magnitude (\textbf{$r$=24.5,} M$_{\text{UV}}$=-21) matches that of the KLCS sample. We note that the $M_{\text{UV}}$ distributions of the two samples are quite different; roughly symmetric about -21 for the KLCS and strongly biased to fainter magnitudes in this work (see Figure \ref{fig:summary_stats}, panels 5 and 6). Additionally, since the HETDEX sample is specifically selected from Ly$\alpha$ emitters (generally young, blue, and compact) there may be a bias for galaxies with enhanced Lyman continuum escape. For consistency with the KLCS, a flux normalization is applied to the resulting \textbf{weighted biweight} stacked spectrum. \textbf{Due to the lower signal-to-noise ratio of the HETDEX data, the normalization is applied to the stack and not individually as in the KLCS case}. This flux normalization of the bright galaxy stack from this work approximates the $f_{1500}$ normalization of the KLCS sample (\textbf{where the KLCS spectra are individually normalized to their flux averaged over 1475-1525~\AA\ prior to stacking}). Since the rest-frame spectral range of the samples from this work does not extend to 1500\AA, a rough translation is employed using the ratio of the normalized flux densities in the KLCS sample over the fairly flat and feature free regions near 1500\AA\ to that near 1300\AA\ (which is covered in the stack from this work) as a scaling factor (specifically, between 1468~\AA\ and 1496~\AA\ to that of 1268~\AA\ to 1296~\AA). The bright galaxy \textbf{weighted biweight} stack is normalized by its own $f_{1300}$ $\times$ the above ratio \textbf{(i.e. 0.58 $\pm$ 0.042 $\mu$Jy $\times$ 1.09)}. While the subsample size is below the threshold where systematics can have a small impact, as the purpose of this comparison is simply to check the overall curve shape and verify key features, no systematics correction was applied. 

\begin{figure}[ht]
    \centering
    \includegraphics[width=0.5\textwidth]{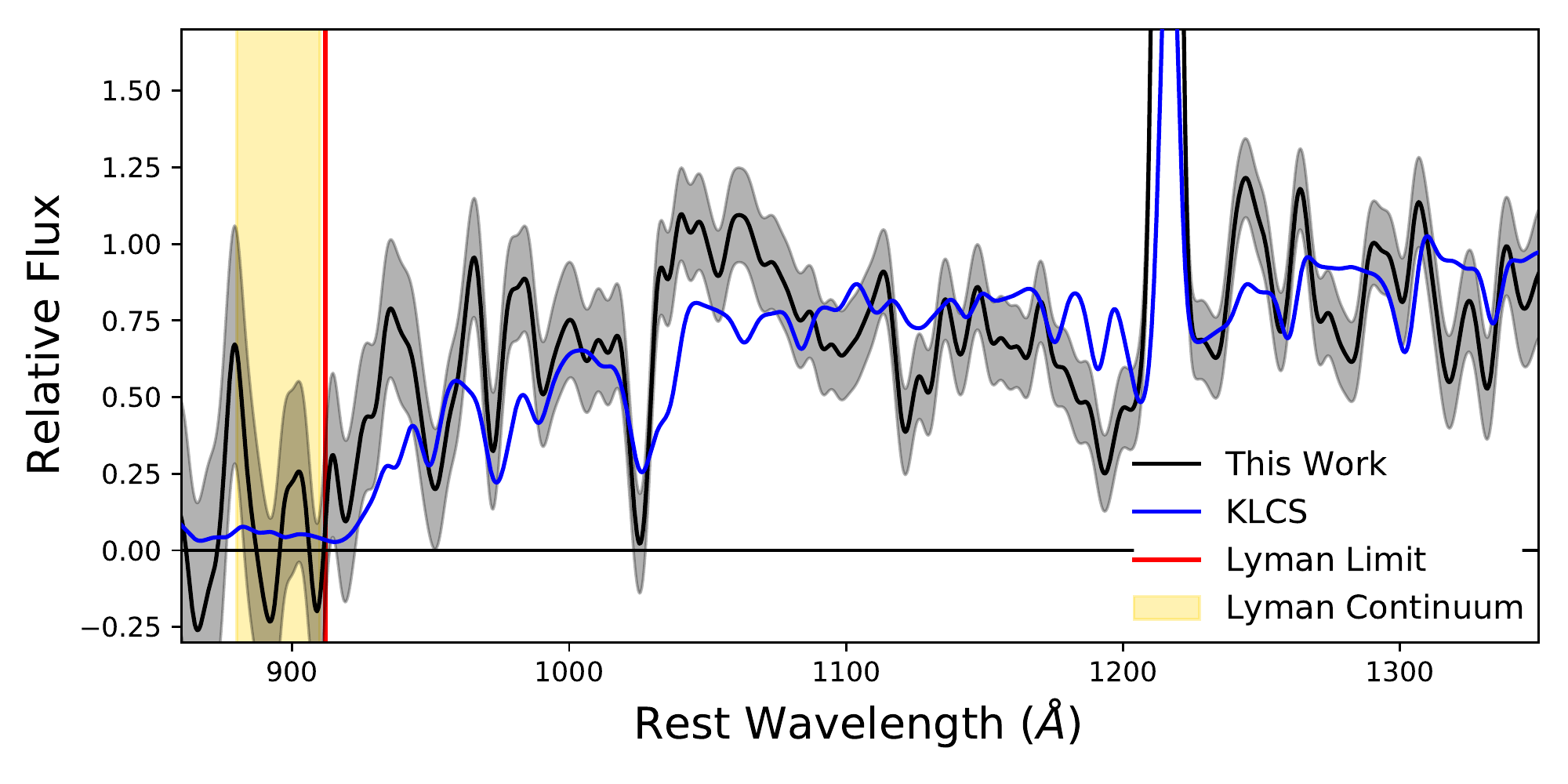}
    \caption{Comparison of a similar subsample (\textbf{52} out of 214) of from this work, based on matching the median \textbf{$r$ and} $M_{\text{UV}}$ to that of \citep{Steidel_2018}. For readability, each spectrum was smoothed with a Gaussian kernel ($\sigma$ = 2.2~\AA). While this subsample is of significantly lower signal-to-noise and spectral resolution compared to KLCS, and the $M_{\text{UV}}$ distribution of the sub-sample in this work skews to fainter magnitudes, the two stacked spectra overlap well, matching up with significant absorption features. See Section \ref{vs_klcs}}
    \label{fig:intflux_vs_steidel}
\end{figure}

This comparison to the KLCS sample is shown in Figure \ref{fig:intflux_vs_steidel}. For readability and to aid in the comparison, the bright subsample (shown in black) and the KLCS sample (shown in blue, without error) are each smoothed with a Gaussian kernel ($\sigma$ = 2.2~\AA\ or 5$\times$ the $~$0.45~\AA\ width of the rest-frame galaxy stack wavelength bins). Although the bright subsample of this work is of a much lower spectral resolution and SNR than the KLCS sample, the two stacks align well and reproduce much the same \textbf{broad} shape and absorption features, lending additional validation to the methodologies of this work. \textbf{Though of limited value given the large noise in this sample of 52, the observed <$f_{900}$ /$f_{1500}$> = 0.07 $\pm$ 0.182, including an IGM correction \citep{Haardt1995,Inoue_2008,Steidel_2018}. While consistent with <$f_{900}$ /$f_{1500}$>$_{out}$ = 0.057 $\pm$ 0.006 in \citet{Steidel_2018}, it is also consistent with zero and a non-detection of Lyman continuum.}

\textbf{
\subsection{Faint Foreground Contamination}}
\textbf{
In an update to \citet{Steidel_2018}, \citet{pahl_2021} uses deep $HST$ imaging to identify several previously undetected, very faint line of sight interlopers contaminating the KLCS sample leading to the removal of 2/15 of the 3$\sigma$ Lyman continuum detections and another 2 from the non-detections (4/124 total). This reduces the full sample ionizing emissivity in the original KLCS findings by $\lesssim10\%$ (from 6.0 to 5.5 $\times 10^{24}$ erg s$^{-1}$ Hz$^{-1}$ Mpc$^{-3}$) and the median Lyman continuum flux density in the 3$\sigma$ detections by $\lesssim5\%$ (from 0.044 to 0.042 $\mu$Jy). \citet{pahl_2021} also performs an alternate ionizing emissivity estimation, retroactively applying it to the \citet{Steidel_2018} sample and finds a decrease of almost 25\% (from 7.2 to 5.5 $\times 10^{24}$ erg s$^{-1}$ Hz$^{-1}$ Mpc$^{-3}$). 
}
\textbf{
While we do not find evidence of significant foreground contamination in our sample, it is a possibility and an issue that we will continue to explore with an expanded HETDEX LAE sample. As a simple check for this work, and using the 3-13\% (4/124 and 2/15) foreground contamination rate in the previous paragraph as a rough guide, we examine the impact of similar contamination on the Lyman continuum detection. We impose a 10\% contamination rate uniformly applied across our entire sample with a worst case scenario of exact line of sight co-location of foreground contaminants with LAE candidates. The contaminants are simulated as simple, flat (in f$_{\nu}$) spectrum with 27 in $g$, assuming brighter foreground galaxies with strong emission lines would be caught by the inspections in Section \ref{sec:selection}. We run 1000 iterations, randomly selecting 10\% of the sample in each run, subtract the simulated contaminant spectrum from those LAE spectra and perform the Lyman continuum stacking analyses. We find that up to 20-25\% of the flux measured in the Lyman continuum region, across all three Methods (Section \ref{sec:analysis} and Table \ref{tab:EUV_subselect_summary}), could come from undetected contamination. As a logical check, we repeat this same test, but remove the randomly impacted LAE candidates under the assumption that the contaminants are now detected. As expected, there is no significant change versus the original weighted biweight measure of the Lyman continuum emission in any of the three Methods and, due to the slightly smaller sample size, the uncertainty in those measures increases by $\sim$5\%. The presence of undetected contamination will have an impact on future work exploring the escaping fraction of EUV photons. Some improvements in contamination detection are anticipated, particularly where deeper, multi-band imaging is available, and more sophisticated simulations will be employed to account for undetected contamination, but this does not change the conclusion of this work of the detection of EUV emission in the ensemble of LAE galaxies.}

\textbf{
\subsection{Biases and Caveats}
}

\textbf{
As noted in the introduction, much of the practical motivation for this work is to provide an alternative survey of z$\sim$3 Lyman continuum leakers that is not subject to the same biases and limitations of other surveys that rely on target pre-selection (typically using the Lyman Break) and deep spectroscopic observations of those targets individually. That is not to say that the methods presented here are free from biases, but the biases are perhaps less significant taken in the context of galaxy ensemble, and some biases, notably in identifying Lyman continuum leakers (described below) are desirable for this work.
}

\textbf{
With the notable exception of the relatively small area MUSE IFU surveys \citep[][and others]{Drake_2017,Feltre_2020}, most other  z$\sim$3 surveys rely on target pre-selection generally using the filter dropout to identify Lyman Break Galaxies (LBG) in the desired redshift range. This is a robust selection method, but does immediately introduce a selection bias that favors galaxies with a strong Lyman Break, i.e. galaxies that are UV bright and Lyman continuum faint, which may tend toward older, more massive galaxies \citep{Gawiser_2007,finkelstein_2010,Jose_2013,santos_2020,santos_2021}. UV faint galaxies, particularly those with stronger Lyman continuum can be missed as they would be faint in the imaging with a weaker Lyman Break. Thus, LBG galaxy surveys may favor galaxies with low EUV emission.  This work’s emission line technique, on the other hand, has no pre-selection and is only flux limited and thus can frequently include fainter galaxies (as noted vs the KLCS above). It does, however, introduce the obvious bias that only galaxies with sufficiently strong Lyman Alpha emission \citepalias{Gebhardt+2021} are included and is blind to non-LAE (or weaker LAE) galaxies. As strong Lyman Alpha emission appears common in Lyman continuum leakers, though strong LyA emission is not sufficient, by itself, to predict Lyman continuum leakage \citep[][]{Kornei_2010,Bian_2020}, there is a potential to favor finding galaxies that are more likely to have stronger EUV emission. This does not appear to be a strong bias as evidenced by the consistent with zero detection of EUV emission in this work’s $M_\text{UV}$-brightest subsample.
}

\textbf{
From the ground, except for extreme cases, z $\sim$ 3 galaxies are generally unresolved point-sources and a slight misalignment of spectrograph singular apertures (a slit or one fiber) with respect to the source centroid can result in an under measurement of flux. Such misalignments can be the result of mechanical positioning errors of the aperture or centering on a centroid that is not co-located with the emission to be studied, though the later should be much less of an issue given the unresolved nature of the sources and sufficiently large apertures. Since HETDEX is a blind survey, no fiber pre-positioning is involved and centroids are found empirically \citepalias{Gebhardt+2021} from the wide mutli-fiber coverage. With the exception of sources that are detected near the edge of an IFU, which are excluded from this work, the loss of light due to fiber placement is minimized. 
}
\textbf{
Many previous pre-selection works stack spectra from a single instrument from data collected over one or a few consecutive (or nearly consecutive) nights and thus have generally similar observing conditions and instrument states. In contrast, though individual HETDEX detections built from three sequential 6-minute exposures, HETDEX stacks are made from observations scattered over many days, seasons and years with varying sky conditions and instrument states. This makes background subtraction considerable more heterogeneous for HETDEX data and, though great care is taken in the background subtraction ( \citetalias{Gebhardt+2021} and this work, section \ref{subsec:background_corr}), it subjects the spectra in this work to larger uncertainties. Fortunately, with the projected very large number of spectra ($\sim$100$\times$ those of this work), the ensemble uncertainties will be minimized.
}

\textbf{
The selection criteria (Section \ref{sec:selection}), which is intended to ensure a contamination free sample, may also introduce some bias in this work. Primarily we are cautious of two criteria. First, the high equivalent width (and other factors) used in the P(Ly$\alpha$) metric (Section \ref{subsec:initial_selection} bullet 4) requirement could introduce a bias against LAEs with weaker Ly$\alpha$ emission and stronger continuum, and thus potentially reduced Lyman continuum leakage, in an effort to avoid contamination from a misidentification of \oii. Second, the restriction to somewhat isolated galaxies (Section \ref{subsec:initial_selection} bullet 6 and Section \ref{subsec:vis_inspection}) could also introduce an environmental bias, though that is unclear as there is no attempt to establish the physical distances between our candidates and their on-sky neighbors. In future datasets, as improvements are made in the classification of HETDEX galaxies \citepalias{Davis2021}, the bias from the P(Ly$\alpha$) selection will be reduced, and added spectral de-blending capabilities will help eliminate unwanted angular/spatial environmental selection effects.\\
}

\section{Summary}

The full-sample $\gtrsim3\sigma$ (\lycujy) detection of Lyman continuum flux (Figure \ref{fig:main_plot}), uncorrected for IGM attenuation, supports the conclusion that stacking the relatively low signal-to-noise, low spectral resolution HETDEX spectra is a valid approach for investigating Lyman continuum photon escape from these Epoch of Reionization analogous galaxies. While the small size of the greatly down-selected dataset in this work restricts meaningful sample subdivision and further immediate investigation, the significance of this detection is encouraging and suggests higher SNR stacks with tight selection criteria for more homogeneous sub-samples are possible.

In the next phase of this investigation, we conservatively expect to increase the 3.0 $< z <$ 3.5 LAE sample size by two orders of magnitude, selecting from a larger initial HETDEX LAE population and by pushing down to lower emission line SNR ($\sim$5.0) and relaxing some of the sub-selection criteria (section $\ref{sec:selection}$) where corrections can be made instead of outright rejection. The much larger sample provides for refined subsampling and higher SNR stacks, enabling meaningful stellar population fitting and the computation of intrinsic EUV escape fractions. This will allow a detailed examination of galaxy properties (UV luminosity, Ly$\alpha$ properties, halo mass, star formation rate, etc) as they relate to Lyman continuum photon production and escape.

Stacking well over 100$\times$ the number of LAE galaxies of other surveys, also affords substantial advantages. By averaging over many sight lines and galaxy orientations, the large individual variations in IGM transmission and potential ISM escape vectors are marginalized out in the aggregation. Additionally, as a blind survey, this data set will not be subject to the same Lyman Break/continuum selection biases and is able, on average, to push into the fainter sources, most similar to typical EoR counterparts \citep[][and others]{Trenti_2010, Finkelstein_2019}. Consequently, we should be able to identify the primary sources of ionizing photons in the EoR and shed additional light on the progression of Reionization. \textbf{Futhermore, this represents a unique and vast dataset for galaxies that serve as laboratories for exploring the physics of the escape of hard ionization radiation.}

\acknowledgments

The authors thank the anonymous referee for the helpful comments which assisted in focusing and improving this manuscript.

HETDEX is led by the University of Texas at Austin McDonald Observatory and Department of Astronomy with participation from the Ludwig-Maximilians-Universit\"at M\"unchen, Max-Planck-Institut f\"ur Extraterrestrische Physik (MPE), Leibniz-Institut f\"ur Astrophysik Potsdam (AIP), Texas A\&M University, The Pennsylvania State University, Institut f\"ur Astrophysik G\"ottingen, The University of Oxford, Max-Planck-Institut f\"ur Astrophysik (MPA), The University of Tokyo, and Missouri University of Science and Technology. In addition to Institutional support, HETDEX is funded by the National Science Foundation (grant AST-0926815), the State of Texas, the US Air Force (AFRL FA9451-04-2-0355), and generous support from private individuals and foundations. We also acknowledge the support from NSF-2008793.

Observations were obtained with the Hobby-Eberly Telescope (HET), which is a joint project of the University of Texas at Austin, the Pennsylvania State University, Ludwig-Maximilians-Universit\"at M\"unchen, and Georg-August-Universit\"at G\"ottingen. The HET is named in honor of its principal benefactors, William P. Hobby and Robert E. Eberly.

The authors acknowledge the Texas Advanced Computing Center (TACC) at The University of Texas at Austin for providing high performance computing, visualization, and storage resources that have contributed to the research results reported within this paper. URL:http://www.tacc.utexas.edu

The Institute for Gravitation and the Cosmos is supported by the Eberly College of Science and the Office of the Senior Vice President for Research at the Pennsylvania State University.


\clearpage


\end{document}